\documentclass[final]{elsart}

\usepackage{bm,amsmath}
\usepackage{algorithm}
\usepackage{algpseudocode}
\usepackage{amsfonts}
\usepackage{graphicx}
\usepackage{enumitem}%
\usepackage{mathrsfs}

\DeclareMathAlphabet\mathpzc{OT1}{pzc}{m}{it}
\let\mathcal=\mathpzc
\def\E{{\mathbb E}}

\def\intinfty{\int\limits_{\!\!-\infty\,\,}^{\,\,\infty\!\!}\kern-0.0em}
\def\iintinfty{\mathop{\int\!\!\int}\limits_{\!\!-\infty\,\,}^{\,\,\infty\!\!}\kern-0.0em}
\def\iiintinfty{\mathop{\int\!\!\int\!\!\int}\limits_{\!\!-\infty\,\,}^{\,\,\infty\!\!}\kern-0.0em}

\let\<=\langle
\let\>=\rangle
\let\^=\hat
\def\~#1{{\-ox{\sf#1}}}
\def\N{{\mathbb N}}
\def\R{{\mathbb R}}
\let\-=\mathbf

\let\-=\mathbf

\def\@#1{{\cal #1}}

\numberwithin{equation}{section}

\journal{Elsevier}

\begin{document}
\centerline{}
\begin{frontmatter}



\title{On an adaptive preconditioned Crank-Nicolson MCMC algorithm for infinite dimensional Bayesian inferences}


\author{Zixi Hu}
\ead{hzx@sjtu.edu.cn} \and
\author{Zhewei Yao}
\ead{zyaosjtu@gmail.com}

\address{Department of Mathematics, Zhiyuan College, Shanghai Jiao Tong University,
Shanghai 200240, China.}

\author{Jinglai Li}
\ead{jinglaili@sjtu.edu.cn}

\address{Institute of Natural Sciences, Department of Mathematics, and MOE Key Laboratory of Scientific and Engineering Computing, Shanghai Jiao Tong University, Shanghai 200240, China. (Corresponding author)}


\medskip
\begin{center}
\end{center}

\begin{abstract}
Many scientific and engineering problems require to perform Bayesian inferences for unknowns of infinite dimension. 
In such problems, many standard  Markov Chain Monte Carlo (MCMC) algorithms become arbitrary slow under the mesh refinement, which is referred to as being dimension dependent. 
To this end, a family of dimensional independent MCMC algorithms, known as the preconditioned Crank-Nicolson (pCN) methods, were proposed to sample the infinite dimensional parameters. 
In this work we develop an adaptive version of the pCN algorithm, where 
the covariance operator of the proposal distribution is adjusted based on sampling history to improve the simulation efficiency.
We show that the proposed algorithm satisfies an important ergodicity condition under some mild assumptions.  
Finally we provide numerical examples to demonstrate the performance of the proposed method.
\end{abstract}


\end{frontmatter}

\section{Introduction}

In many real-world inverse problems,  the unknowns that one wants to estimate are functions of space and/or time. 
Solving such problems with the Bayesian approaches~\cite{kaipio2005statistical,stuart2010inverse}, often require to perform Markov Chain Monte Carlo (MCMC) simulations in function spaces.
Namely one first represents the unknown function with a finite-dimensional parametrization, for example, by discretizing the function on a pre-determined mesh grid, and then performs MCMC simulations in the resulting finite dimensional space. 
It has been known that standard MCMC algorithms, such as the random walk Metropolis-Hastings (RWMH),  can become arbitrarily slow as the discretization mesh of the unknown is refined~\cite{roberts1997weak,roberts2001optimal,beskos2009optimal,mattingly2012diffusion}. 
That is, the mixing time of an algorithm can increase to infinity as the dimension of the discretized parameter approaches to infinity,
and in this case the algorithm is said to be \emph{dimension-dependent}.
To this end, a very interesting line of research is to develop \emph{dimension-independent} MCMC algorithms 
by requiring the algorithms to be well-defined in the function spaces.
In particular, a family of dimension-independent MCMC algorithms, known as the preconditioned Crank Nicolson (pCN) algorithms, were presented in \cite{cotter2013mcmc} by constructing a Crank-Nicolson 
discretization of a stochastic partial differential equation (SPDE) that preserves the reference measure. 

The sampling efficiency of the pCN algorithm can be improved by incorporating the data information in the proposal design,
and a popular way to achieve this goal is the adaptive MCMC methods. 
Simply speaking, the adaptive MCMC algorithms improve the proposal based on the sampling history from the targeting distribution~(c.f. \cite{andrieu2008tutorial,atchade2009adaptive,roberts2009examples} and the references therein) as the iterations proceed. A major advantage of the adaptive methods is that they only require the ability
to evaluate the likelihood functions, which makes them particularly convenient for problems 
with black-box models. 
In a recent work~\cite{feng2015adaptive}, we develop an adaptive independence sampler MCMC algorithm for the infinite dimensional problems. 
A main difficulty of independence sampler MCMC algorithms is that the efficiency of such algorithms depends critically on the ability of the chosen proposal, often in a parametrized form,
 to approximate the posterior in the entire state space, and the algorithm may perform very poorly if the proposal  
can not well approximate the posterior distribution. In this respect, random walk based algorithms may be more convenient to use,
 as they do not require such a ``global proposal''.
In this work, we present an adaptive random walk MCMC based on the preconditioned Crank-Nicolson (pCN) algorithm in  \cite{cotter2013mcmc}.
 Specifically, we adaptively adjust the covariance operator of the proposal to improve the sampling efficiency. 
We parametrize the covariance operator in a specific form that has been used in \cite{pinski2015algorithms,feng2015adaptive}, 
and we provide an algorithm that can efficiently update the parameter values as the iteration proceeds. 
By design, the acceptance probability of our algorithm is well defined
and thus the algorithm is dimension independent.
Moreover, we can show that the algorithm satisfies some important ergodicity conditions in the infinite dimensional setting. 
Note that, another existing adaptive MCMC algorithm for infinite dimensional problems is the dimension independent adaptive Metropolis (DIAM) proposed in \cite{chen2015accelerated}.
The DIAM is also based on the pCN algorithm, 
but our method preserves an important feature of the standard pCN algorithm, i.e., 
the acceptance probability being independent on the proposal distribution, while the DIAM method does not.

We note that, an alternative class of methods improve the sampling efficiency by guiding the proposal with the local derivative information of the likelihood function.
Such derivative based methods include: the stochastic Newton MCMC~\cite{martin2012stochastic,petra2014computational},
the  Riemann manifold Hamiltonian MC~\cite{bui2014solving},
the operator-weighted proposal method~\cite{law2014proposals},  the dimension-independent likelihood-informed MCMC~\cite{cui2016dimension}, 
 the generalized pCN algorithm~\cite{rudolf2015generalization}, and so on.  
We reinstate that in this work we are focused on the type of problems where the derivative information is difficult to obtain, 
and thus those derivative based methods are not in our scope.

The rest of the paper is organized as the following. In section~\ref{sec:method} we describe the setup of infinite dimensional inference problems and present our adaptive  algorithm in detail.  
In section~\ref{sec:examples} we provide several numerical examples to demonstrate the performance of the proposed algorithm. 
Finally we offer some concluding remarks in section~\ref{sec:conclusion}.

\section{The adaptive pCN algorithm}\label{sec:method}

\subsection{Bayesian inferences in function spaces}
We present the standard setup of the Bayesian inverse problem following \cite{stuart2010inverse}. 
We consider a separable {Hilbert} space $X$ with inner product $\<\cdot,\cdot\>_X$.
 Our goal is to estimate the unknown  $u\in X$ from data $y\in Y$ where $Y$ is the data space and $y$ is related to $u$ via a likelihood function $\exp(-\Phi^y(u))$.
In the Bayesian inference we assume that the prior $\mu_0$ of $u$, is a  (without loss of generality)~zero-mean Gaussian measure defined on $X$ with covariance operator $\@C_0$,
i.e. $\mu_0 = N(0,\@C_0)$. 
Note that $\@C_0$ is symmetric positive and of trace class.
The range of $\@C_0^{\frac12}$,
\[E = \{u = \@C_0^{\frac12} x\, |\, x\in X\}\subset X,\]
which is a Hilbert space equipped with inner product~\cite{da2006introduction},
\[\<\cdot,\cdot\>_E = \<\@C_0^{-\frac12}\cdot,\@C_0^{-\frac12}\cdot\>_X ,\]
is called the Cameron-Martin space of measure $\mu_0$. 
In this setting, the posterior measure $\mu^y$ of $u$ conditional on data $y$
is provided by the Radon-Nikodym derivative:
\begin{equation} \frac{d\mu^y}{d\mu_{0}}(u) =\frac1Z\exp(-\Phi^y(u)),
 \label{e:bayes}
\end{equation}
with $Z$ being a normalization constant, 
which can be interpreted as the Bayes' rule in the infinite dimensional setting.
In what follows, without causing any ambiguity, we shall drop the superscript $y$ in $\Phi^y$ and $\mu^y$ for simplicity, while keeping 
in mind that these items depend on the data $y$.  
For the inference problem to be well-posed, one typically requires the functional $\Phi$ to satisfy the Assumptions (6.1) in \cite{cotter2013mcmc}.
It is known that there exists a complete orthonormal 
basis $\{e_j\}_{j\in\N}$ on $X$ and a sequence of non-negative numbers $\{\alpha_j\}_{j\in\N}$
such that ${\@C_0} e_j = \alpha_j e_j$ and $\sum_{j=1}^\infty \alpha_j <\infty$, i.e., 
 $\{e_j\}_{k\in\N}$ and $\{\alpha_j\}_{k\in\N}$ being the eigenfunctions and eigenvalues of $\@C_0$ respectively~(\cite{da2006introduction}, Chapter~1).
For convenience's sake, we assume that the eigenvalues are in a descending order: $\alpha_j\geq\alpha_{j+1}$ for any $j\in\N$.
$\{e_j\}_{j=1}^\infty$ are known as the Karhunen-Lo\`eve~(KL) modes associated with $\@N(0,\@C_0)$.

\subsection{The Crank-Nicolson algorithms}
We start by briefly reviewing the family of Crank-Nicolson (CN) algorithms for infinite dimensional Bayesian inferences, developed in \cite{cotter2013mcmc}. 
Simply speaking the algorithms are based on the stochastic partial differential equation (SPDE)
\begin{equation}
\frac{du}{ds}=-\mathcal{K}\mathcal{L}u+\sqrt{2\mathcal{K}}\frac{db}{ds},\label{e:spde}
\end{equation}
where $\@L=\@C_0^{-1}$ is the precision operator for $\mu_0$, $\@K$ is a positive operator,
 and $b$ is a Brownian motion in $X$ with covariance operator the identity. 
The proposal is then derived by applying the CN discretization to the SPDE~\eqref{e:spde}, yielding, 
\begin{equation}
v=u-\frac12\delta\mathcal{K}\mathcal{L}(u+v)+\sqrt{2\mathcal{K}\delta}\xi_0, \label{e:prop}
\end{equation}
for a white noise $\xi_0$ and $\delta\in(0,2)$.
In \cite{cotter2013mcmc}, two choices of $\@K$ are proposed, resulting in two different algorithms.
First, one can choose $\@K=\@I$, the identity, obtaining: 
\[
(2\@C+\delta \@I)v=(2\@C-\delta \@I)u+\sqrt{8\delta}w,
\]
where $w\sim \@N(0,\@C_0)$,
which is known as the plain CN algorithm. 
Alternatively one can choose $\@K=\@C_0$, resulting in the pCN proposal:
\begin{equation}
v=(1-\beta^2)^{\frac12}u+ \beta w, \label{e:pcn}
\end{equation}
where \[\beta = \frac{\sqrt{8\delta}}{2+\delta}.\]
It is easy to see that $\beta\in[0,1]$.
In both CN and pCN algorithms,  the acceptance probability is
\begin{equation}
a(v,u) = \min\{1, \exp{\Phi(u)-\Phi(v)}\}. \label{e:acc}
\end{equation}

\subsection{The adaptive algorithm}\label{pr:acc}
To derive the new algorithm, we rewrite the proposal Eq.~\eqref{e:prop} as
\begin{equation}
v=\frac{(I-\frac12\delta\mathcal{K}\mathcal{L})}{(I+\frac12\delta\@K\@L)}u
+\frac{\sqrt{2\delta\@K}}{(I+\frac12\delta\mathcal{K}\mathcal{L})}\xi_0, \label{e:prop2}
\end{equation}
Now we do a substitution. Namely we let 
\begin{equation}
\frac{\sqrt{2\delta\@K}}{(I+\frac12\delta\mathcal{K}\mathcal{L})} = \beta \sqrt{\@B}, \label{e:sub1}
 \end{equation}
and by some simply calculation, we can verify that
\begin{equation}
\frac{(I-\frac12\delta\mathcal{K}\mathcal{L})}{(I+\frac12\delta\@K\@L)} = \sqrt{(\@I-\beta^2 \@B\@L)}.\label{e:sub2}
\end{equation}
Substitute Eqs.~\eqref{e:sub1} and \eqref{e:sub2} into Eq.~\eqref{e:prop2}, and we obtain a new proposal:
\begin{equation}
v = {(\@I-\beta^2 \@B\@L)^{\frac12}}u+ \beta w \label{e:propnew}
\end{equation}
where $w\sim \@N(0,\@B)$.
This proposal can be understood as a special case of the generalized pCN or the operator weighted proposal. 
The major difference is that in those two methods, the operator is determined by the derivative information of the likelihood function, while in our algorithm,
it is determined with an adaptive method. Before discussing the details of how to determine the operator $\@B$, we first show that under mild conditions, the proposal~\eqref{e:propnew} results in well-defined acceptance probability in a function space:
\begin{prop}
Suppose operator $\@B$ is symmetric positive and of trace class. Let $q(u,\cdot)$ be the proposal distribution associated to Eq.~\eqref{e:propnew}. 
Define measures $\eta(du,dv)=q(u,dv)\mu(du)$ and $\eta^\bot(du,dv)=q(v,du)\mu(dv)$ on $X\times X$. If $\@B$ commutes with $\@C_0$,  $\eta^\bot$ is absolutely continuous with respect to $\eta$, and 
\[
\frac{d\eta^\bot}{d\eta}(u,v)= \exp(\Phi(u)-\Phi(v)).
\]
\end{prop}
\begin{pf}
Define $\eta_0(du,dv)=q(u,dv)\mu_0(du)$. The measure $\eta_0$ is Gaussian. From $\@B$ and $\@C_0$ are commutable, we have
\[
\E^{\eta_0}v\otimes v=(\@I-\beta^2\@B\mathcal{L})\@C_0+\beta^2\@B=\@C_0=\E^{\eta_0}u\otimes u.
\]
Then $\eta_0$ is symmetric in $u,v$. Now
\[
\eta(du,dv)=q(u,dv)\mu(du),\quad
\eta_0(du,dv)=q(u,dv)\mu_0(du),
\]
and $\mu$,$\mu_0$ are equivalent. It follows that $\eta$ and $\eta_0$ are equivalent and 
\[
\frac{d\eta}{d\eta_0}(u,v)=\frac{d\mu}{d\mu_0}(u)=  \frac1{Z}\exp(-\Phi(u)).
\]
Since $\eta_0$ is symmetric in $u,v$ we also have that $\eta^\bot$ and $\eta_0$ are equivalent and that 
\[
\frac{d\eta^\bot}{d\eta_0}(u,v)= \frac1Z\exp(-\Phi(v)).
\] 
Since equivalence of measures is transitive if follows that $\eta$ and $\eta^\bot$ are equivalent and 
\[
\frac{d\eta^\bot}{d\eta}(u,v)= \exp[{\Phi(u)}-{\Phi(v)}].
\]
\end{pf}

It follows immediately from the detailed balance condition that the associated acceptance probability of proposal~\eqref{e:propnew} is also given by Eq.~\eqref{e:acc}.

Now we discuss how to specify the operator $\@B$, and we start with assuming $\@B$ an appropriate parametrized form. Note that an essential condition in Proposition~\ref{pr:acc} is that $\@B$ must commute with
$\@C_0$. To satisfy this condition, it is convenient to design a $\@B$ that has common eigenfunctions with $\@C_0$. 
Namely, we write $\@B$ in the form of
\begin{equation} \label{e:B}
\@B \,\cdot = \sum_{j=1}^\infty \lambda_{j}\<e_j,\cdot\>e_j,
\end{equation}
 with $\lambda_{j}$ being the coefficients. 
It is easy to see that $\@B$ is a symmetric operator with 
eigenvalue-eigenfunction pair $\{\lambda_j,e_j\}_{j=1}^\infty$,
which implies that $\@B$ and $\@C_0$ commute.

A well-adopted rule in designing efficient MCMC algorithms is that the proposal covariance should be close to the
covariance operator of the posterior~\cite{roberts2001optimal,haario2001adaptive}. 
Now suppose the posterior covariance is $\@C$, and
one can determine the proposal covariance $\@B$ by solving
  \begin{equation}
  \min_{\{\lambda_j\}_{\lambda=1}^\infty}\|\@B-\@C\|_{HS}, \label{e:minhs}
  \end{equation}
   where $\|\cdot\|_{HS}$ is the Hilbert-Schmidt norm defined as $\|\@A\|^2_{HS}=\mathrm{Tr}(\@A^* \@A)$ where $\@A$ is any bounded operator on $X$ and
	$\@A^*$ is the adjoint of $\@A$.
 By some basic algebra, we can show that the optimal solution of Eq~\eqref{e:minhs} is
 \[\lambda_j = \< \@C e_j,e_j\>^{-1}\]
 for $j=1...\infty$.
Since $\@C$ is the posterior covariance,  for any $v$ and $v' \in X$, we have~\cite{da2006introduction},
 \begin{equation}
\<\@C v, v'\> =\int \<v,u-m\>\<v',u-m\> \mu(du), \label{e:cov}
\end{equation}
where $m$ is the mean of $\mu$.
Using Eq.~\eqref{e:cov}, we can derive that
\begin{equation}
 \lambda_j =\int(x_{j}-u_j)^2\mu(du), \label{e:lambda_j}
\end{equation}
where $x_j=\<m,e_j\>$ and $u_j = \<u,e_j\>$ for $j = 1...\infty$.

In practice, the posterior covariance $\@C$ is not directly available, and so here we determine the operator $\@B$ with an adaptive MCMC algorithm. 
Simply speaking, the adaptive algorithm starts with an initial guess of $\@B$ and then adaptively updates the $\@B$ based on the sample history.
Estimating all eigenvalues from the sample history is not practical due to the finite sample size. Here we make a finite-dimension reduction:
namely, only the first $J$ eigenvalues are given in the form of Eq.~\eqref{e:lambda_j} which is further estimated from the sample history and the rest of them are taken to be fixed. 
In particular we let 
\begin{equation} \label{e:lambda2}
\lambda_j= \begin{cases} \int(x_{j}-u_j)^2\mu(du) &\mbox{for } j \leq J \\
\alpha_j & \mbox{for } j >J. \end{cases}  
\end{equation}
The argument that we compute $\lambda_j$ as is in Eq.~\eqref{e:lambda2} may become 
 more clear if  we look at the projections of the proposal onto each eigenmodes:
\begin{equation}\label{e:projection}
\<v,e_j\> =\begin{cases} (1-\beta^2 \lambda_i/\alpha_i )^{\frac12}u_j+\beta w_j \quad \mbox{where} \,\,w_j\sim \@N(0,\lambda_j) &\mbox{for } j \leq J \\
(1-\beta^2)^{\frac12}u_j+\beta w_j \quad \mbox{where} \,\,w_j\sim \@N(0,\alpha_j)& \mbox{for } j >J. \end{cases}
\end{equation}
Eq.~\eqref{e:projection} shows the basic scheme of the algorithm:
it performs an adaptive pCN for the KL modes $j\leq J$ with the proposal covariance adapted 
to approximate that of the posterior,
and a standard pCN for all $j>J$.
The intuition behind our algorithm is based on the assumption that the (finite-resolution) data is only informative about a finite number 
of KL modes of the prior.  In particular, the data can not provide information about the modes that are highly oscillating (associated with small eigenvalues) and for those modes, the posterior is 
approximately the prior. 
In this case, for the modes that are informed by the data, we shall adjust the eigenvalues to approximate the posterior covariance; 
for those that are not, the best strategy is to simply use the covariance of the prior (which is also the posterior). 

Now we discuss how to update the values of $\lambda_j$ from posterior samples for $i=1...J$. 
To this end, suppose we have a set of posterior samples $\{u^n\}_{i=0}^n$, and 
the values of parameters $\lambda_j$ are estimated using  the sample average approximation of Eq.~\eqref{e:lambda_j}:
\begin{subequations}
\label{e:h_j2}
\begin{gather}
 x^n_{j} = \frac1{n+1}\sum_{i=0}^n \<u^{i},e_j\>,\\
s^{n}_j = \sum_{i=0}^n (u_j^{n})^2,\\
\lambda^n_j ={\frac1{n+1} \sum_{i=0}^n(x^n_{j}-u^i_j)^2+\epsilon^2}, 
 \end{gather}
\end{subequations}
for $j=1...J$. 
Here $\epsilon$ is a small constant, introduced to ensure the stability of the algorithm, i.e., to keep  $\lambda^n_j$ from becoming arbitrarily small.
For efficiency's sake, we can rewrite Eq~\eqref{e:h_j2} in a recursive form
\begin{subequations}
\label{e:h_j3}
\begin{gather}
 x^{n}_{j} = \frac{n}{n+1}x_j^{n-1}+ \frac1{n+1} \<u^{n},e_j\>,\\
 s^{n}_j = s_j^{n-1}+(u_j^{n})^2,\\
 \lambda_j^{n} ={\frac1{n+1} s_j^n-(x_j^n)^2} , 
 \end{gather}
\end{subequations}
for $j=1...J$ and $n>0$.
Note here that, in principle the estimated $\lambda_j^n$ from samples can be arbitrarily large, which causes issues as $(\@I-\beta^2\@B\@L)$ must not be negative. 
Thus we let $\lambda^n_j = \min\{\lambda^n_j,\alpha_j\}$ for $j=1...J$, and as a result $\lambda_j\leq \alpha_j$ for $j=1...J$. 
It is easy to see that the operator $\@B$ resulting from $\{\lambda_j^n\}_{j=1}^J$ 
 is symmetric positive and of trace class. 
Finally we note that, it is not robust to estimate the parameter values with a very small number of samples, 
and to address the issue, we first draw a certain number of samples with a standard pCN algorithm before starting the adaptation.  
We describe the complete  adaptive pCN (ApCN) algorithm in Algorithm~\ref{al:apcn}. 
\begin{algorithm}[!tb]
 \caption{The adaptive pCN algorithm}
    \label{al:apcn}
    \begin{algorithmic}[1]
    
 \State  Initialize $u^0\in S $;
\For {$n=0$ to $N'$}

	\State Propose $v$ using Eq~\eqref{e:pcn};	
        \State Draw $\theta\sim U[0,1]$
        \State Let $a: = \min\{1, \exp[{\Phi(u^n)}-{\Phi(v)}]\}$;
        
        \If {$\theta\leq a$ } 
            \State $u^{n+1}=v$;
						\Else
						\State $u^{n+1}=u^{n}$;
        \EndIf
			
				    \EndFor
\State Compute $\{x_j^{N'},s_j^{N'},\lambda_j^{N'}\}_{j=1}^J$ using Eq.~\eqref{e:h_j2} and samples $\{u^i\}_{i=1}^{N'}$;
\For{$j=1$ to $J$} 
\State $\lambda_j=\min\{\lambda_j,\alpha_j\}$;\EndFor
   \For {$n=N'$ to $N$}
       \State Compute $\@B$ from Eqs.~\eqref{e:B} with $\{\lambda_j^{n}\}_{j=1}^J$;
	\State Propose $v$ using Eq~\eqref{e:propnew};	
        \State Draw $\theta\sim U[0,1]$
        \State Let $a: = \min\{1, \exp[{\Phi(u^n)}-{\Phi(v)}]\}$;
        
        \If {$\theta\leq a$ } 
            \State $u^{n+1}=v$;
						\Else
						\State $u^{n+1}=u^{n}$;
        \EndIf
				\State Compute $\{x_j^{n+1},s_j^{n+1},\lambda_j^{n+1}\}_{j=1}^J$ using Eqs.~\eqref{e:h_j3};
				    \EndFor

    \end{algorithmic}
		\medskip
		
    \end{algorithm}

Finally an important issue in the implementation is to determine the number of adapted eigenvalues $J$.
Here we propose to let $J =\min\{j\in \N\}$ such that, 
\[\frac{\sum_{i=1}^j{\alpha_j}}{\sum_{i=1}^\infty{\alpha_i}}>\rho,\]
where $0<\rho<1$ is a prescribed number (e.g. $\rho=0.99$).

\subsection{Ergodicity analysis}

It is well known that, the adaptation may destroy the ergodicity of the algorithm,
and as a result the chain constructed may not converge to the target distribution.  
It has been suggested by Roberts and Rosenthal \cite{roberts2009examples} that, an adaptive MCMC algorithm
has the correct asymptotic convergence, provided that it satisfies 
the Diminishing  Adaptation~(DA) condition, which, loosely speaking, 
requires the transition probabilities to converge as the iteration proceeds,  
 and the Containment condition.
As the latter is regarded as merely a technical condition which is satisfied for virtually all reasonable adaptive schemes~\cite{roberts2009examples},
it often suffices to prove an adaptive algorithm satisfies the DA condition. 
Next we show that  the proposed ApCN algorithm satisfies  the DA condition under a minor modification. 
Namely, we change Eq.~\eqref{e:bayes} to be
\begin{equation} \frac{d\mu^y}{d\mu_{0}}(u) = 
\begin{cases}
\frac1Z\exp(-\Phi(z)), &\|u\|_X\leq R\cr
0,&\|u\|_X> R,
\end{cases}
\label{e:lh_mod}
\end{equation}
where $R$ is a prescribed positive constant. 
We want emphasize here that, just like the work~\cite{haario2001adaptive}, the purpose of the modification is to simplify our proof here,
and practically speaking, its impact on the inference results should be negligible, provided that $R$ is taken to be sufficiently large.  

Let us now set up some notations. 
Assume that $\@B_n(u_0,u_1,\cdots,u_{n-2},u)$ is the operator $\@B$ at iteration $n$ computed with $u_0, u_1, \cdots, u_{n-2}, u$ through Algorithm 1. For simplicity, we define 
\[
\@B_{n,\zeta_{n-2}}(u)=\@B_n(u_0,u_1,\cdots,u_{n-2},u), \quad \mathrm{where}\quad\zeta_{n-2}=(u_0,u_1,\cdots,u_{n-2}),
\]
and let $\{\lambda_{n,i}\}_{i=1}^\infty$ be the eigenvalues of $\@B_{n,\zeta_{n-2}}(u)$. 
We define $q_{n,\zeta_{n-2}}(u;dv)$ to be the proposal distribution 
associated to
\[
v = {(\@I-\beta^2 \@B_{n}(\zeta_{n-2},u)\@L)^{\frac12}}u+ \beta w 
\]
where $w\sim N(0,\@B_{n,\zeta_{n-2}}(u))$, and
\[
Q_{n,\zeta_{n-2}}(u,dv)=a(u,v)q_{n,\zeta_{n-2}}(u,dv)+\delta_u(dv)(1-\int a(u,z)q_{n,\zeta_{n-2}}(u,dz))
\] 
where $a(\cdot,\cdot)$ is given by Eq.~\eqref{e:acc}.
It can be verified that
\begin{equation}
\sigma_{n,i}=(1-\beta^2\frac{\lambda_{n,i}}{\alpha_{i}})^{\frac12},
\label{e:sigma}
\end{equation}
are the eigenvalues of $(\@I-\beta^2 \@B_{n,\zeta_{n-2}}(u)\@L)^{\frac12}$.
We then have the following theorem: 
\begin{thm}[DA condition] \label{th:da}
 There is a fixed positive constant $\gamma$ such that 
\[
\sup_{u\in X}\|Q_{n,\zeta_{n-2}}(u,\cdot)-Q_{n+1,\zeta_{n-1}}(u,\cdot)\|\leq\frac{\gamma}{n}
\]
for any $\zeta_{n-1}$ and $\zeta_{n-2}$ such that $\zeta_{n-1}$ is a direct continuation of $\zeta_{n-2}$.  
Here  $\|\cdot\|$ is the total variation norm.
\end{thm}
We provide the proof of the theorem in the Appendix. 
   
\section{Numerical examples}\label{sec:examples}
\subsection{An ODE example}  Our first example is a simple inverse problem where the forward model is governed by an ordinary differential equation (ODE):
\[\frac{\partial x(t)}{\partial t} = -u(t)x(t) \]
with a prescribed initial condition. Suppose that we observe the solution $x(t)$ several times in the interval $[0,T]$, and we want to infer the unknown coefficient 
$u(t)$ from the observed data.

\begin{figure}
\centerline{\includegraphics[width=.49\textwidth]{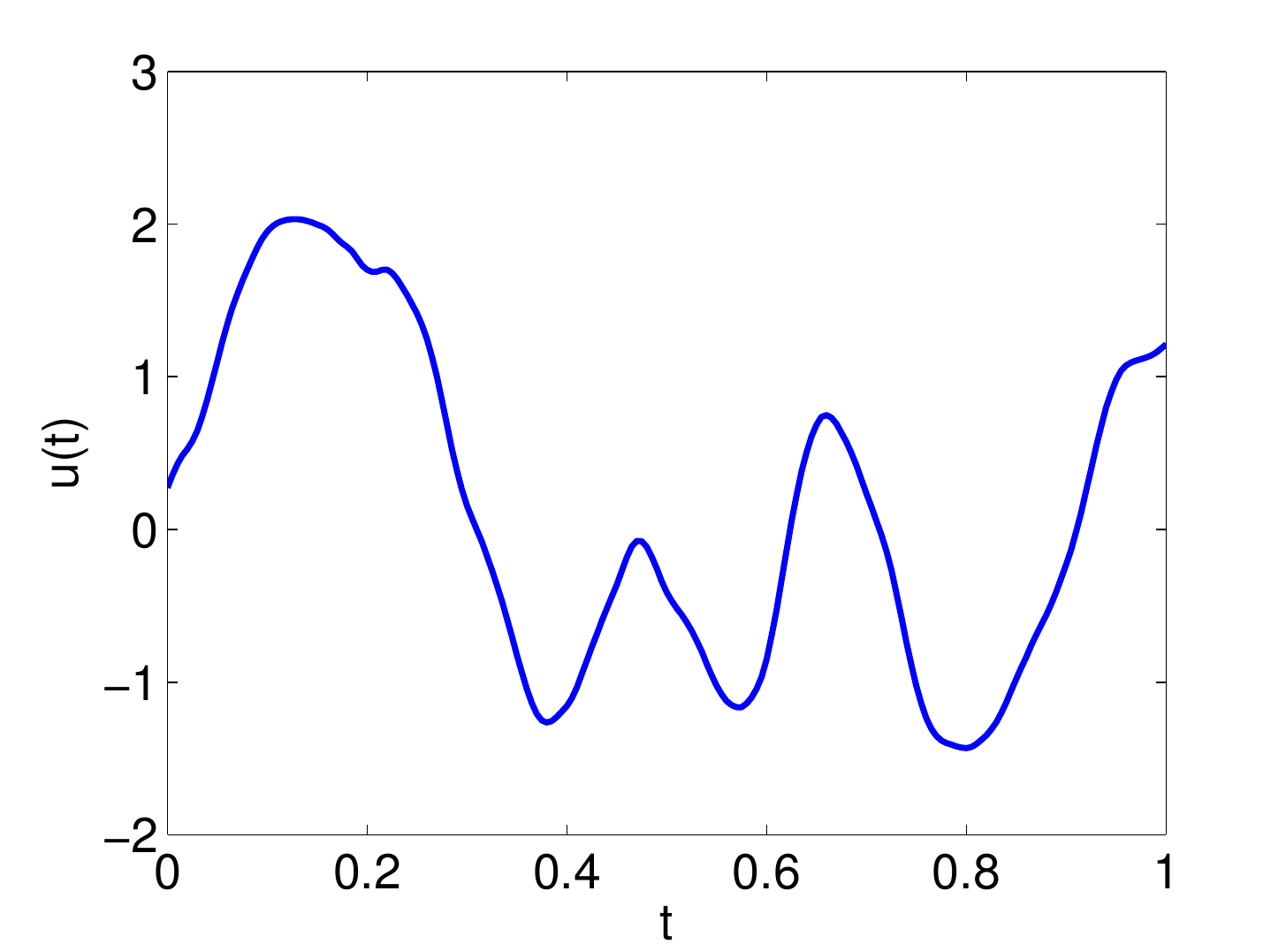}
\includegraphics[width=.49\textwidth]{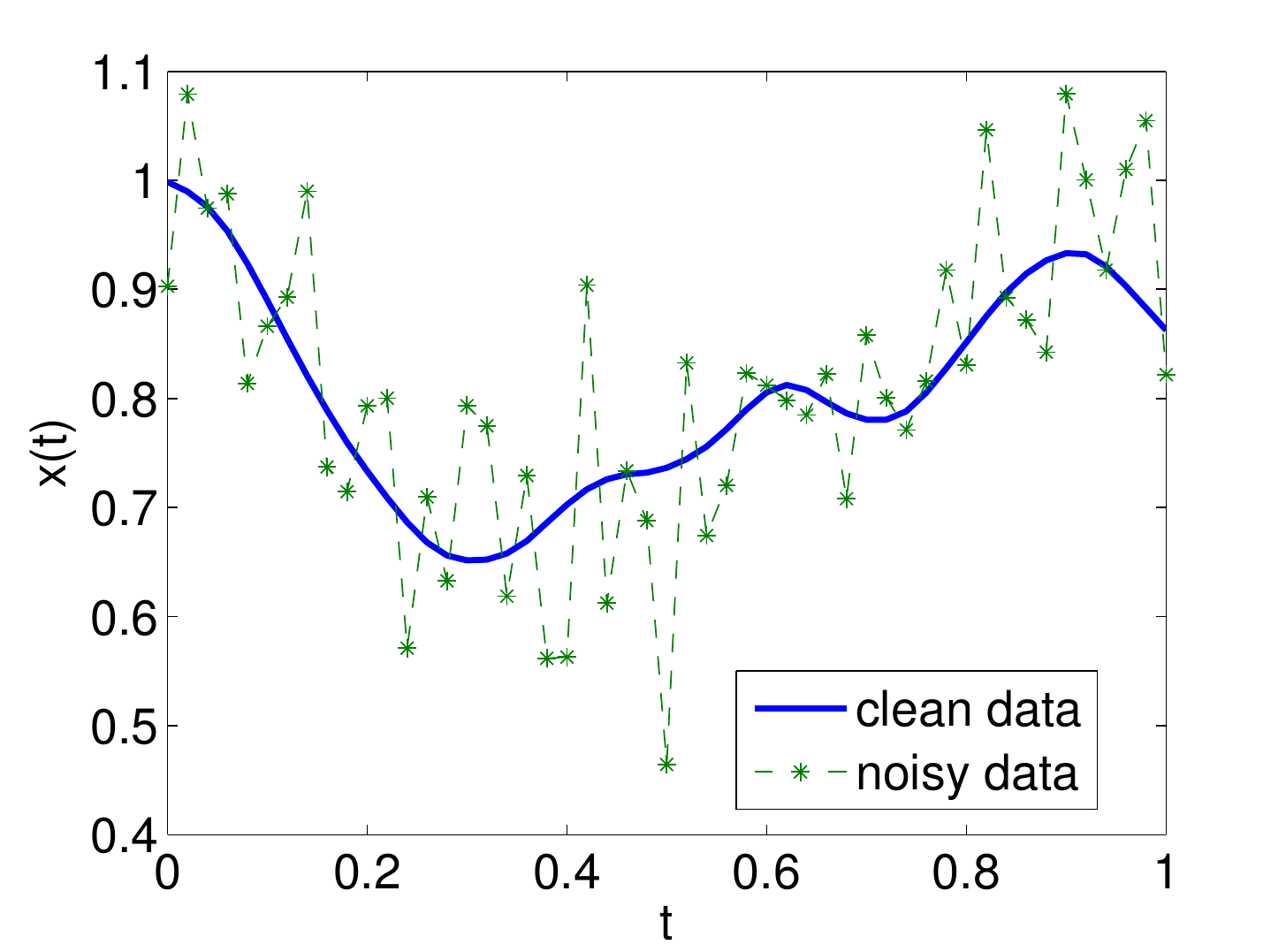}}
\caption{(for the ODE example) The truth (Left) and the data simulated with it (Right).}
\label{f:data_ode}
\end{figure}

In our numerical experiments, we let the initial condition be $x(0) = 1$ and $T = 1$. Now suppose that the solution is measured every $T/100$ time unit from $0$ to $T$ and the error in each measurement is assumed to be an independent Gaussian $N(0,0.1^2)$. 
The prior is taken to be a zero mean Gaussian with Mat\'ern covariance~\cite{rasmussen2006gaussian}:
\[
K(t_1,t_2) = \sigma^2 \frac{2^{1-\nu}}{\Gamma(\nu)}(\sqrt{2\nu}\frac{d}{l})^\nu B_\nu(\sqrt{2\nu}\frac{d}{l}),
\]
where $d=|t_1-t_2|$, $\Gamma(\cdot)$ is the Gamma function, and $B_\nu(\cdot)$ is the modified Bessel function. 
A random function with the Mat\'ern covariance is $[\nu-1]$ mean square (MS) differentiable. 
Several authors suggest that the Mat\'ern covariances can often provide a better model for many real-world physical processes than the popular squared exponential covariances~\cite{rasmussen2006gaussian}. 
In this example, we choose  $l=1$, $\sigma=1$, and $\nu=5$ implying second order MS differentiability.
In the numerical tests, we represent the unknown with $501$ grid points.
We use synthetic data that is 
 generated by applying the forward model to a true coefficient $u$ and then adding noise to the result. The true coefficient is randomly drawn from the prior distribution. 
Both the truth and the simulated data are shown in Fig.~\ref{f:data_ode}.
We perform the proposed adaptive pCN algorithm with $1\times10^6$ samples and another $5\times10^4$ pCN samples are used in the pre-run. 
We set the stepsize $\beta = 1/5$,
 and we choose $\rho=0.99$ resulting in $J=14$, i.e., $14$ eigenvalues being adapted.  

\begin{figure}
\centerline{\includegraphics[width=.49\textwidth]{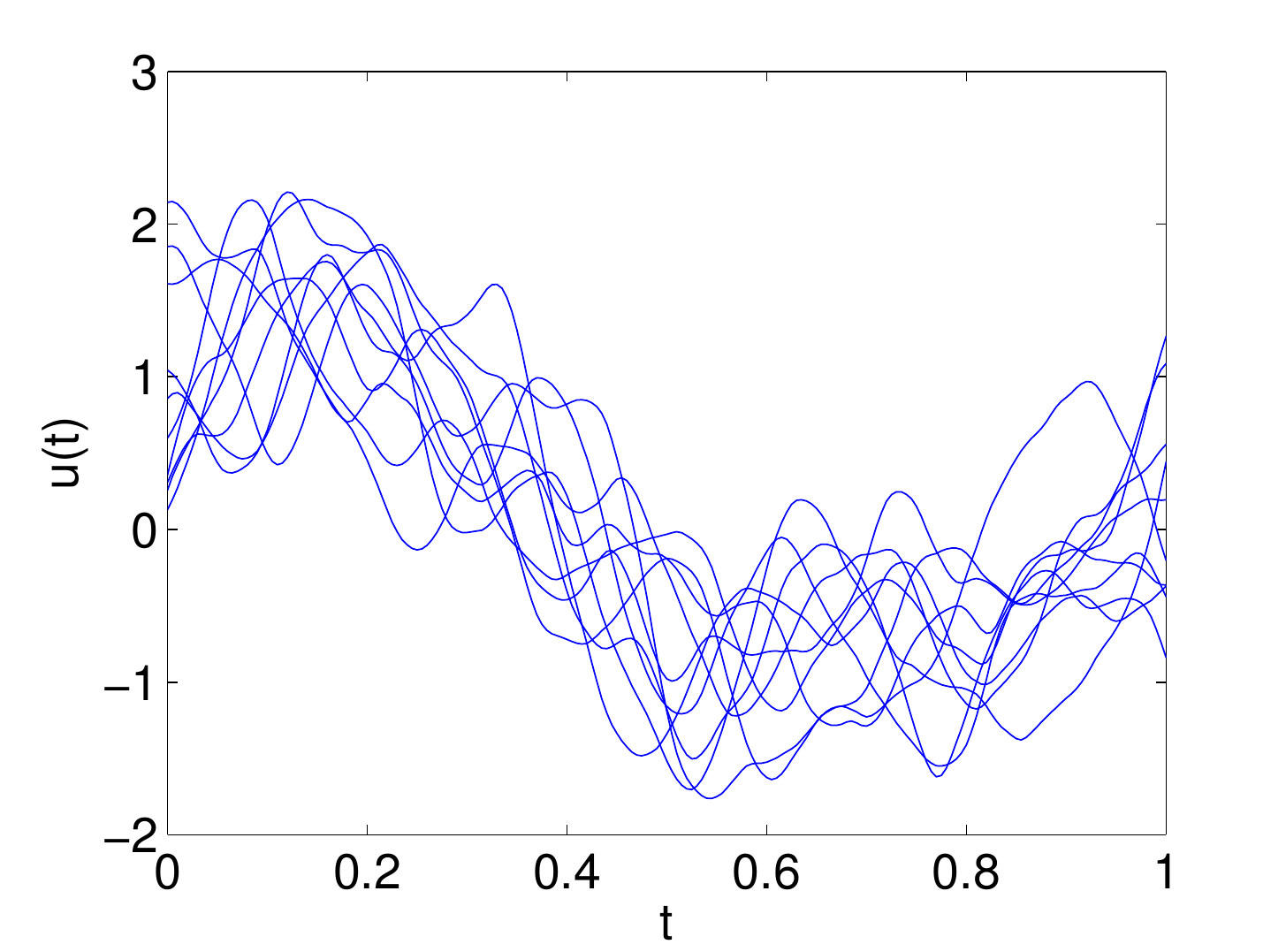}
\includegraphics[width=.49\textwidth]{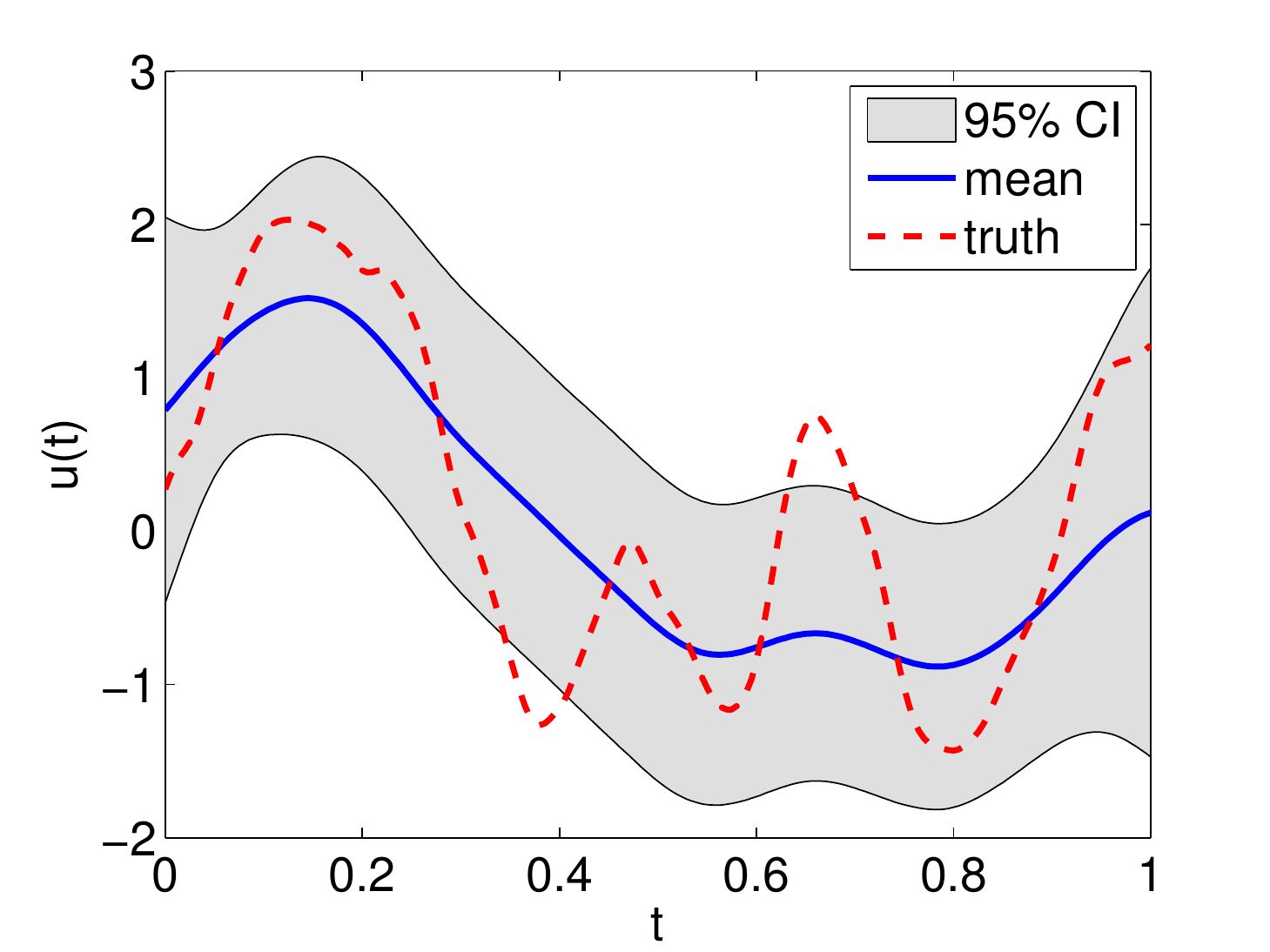}}
\caption{(for the ODE example) Left: 10 randomly drawn samples from the posterior. Right: the posterior mean and the $95\%$ confidence interval.}
\label{f:mean_ode}
\end{figure}

\begin{figure}
\centerline{\includegraphics[width=.49\textwidth]{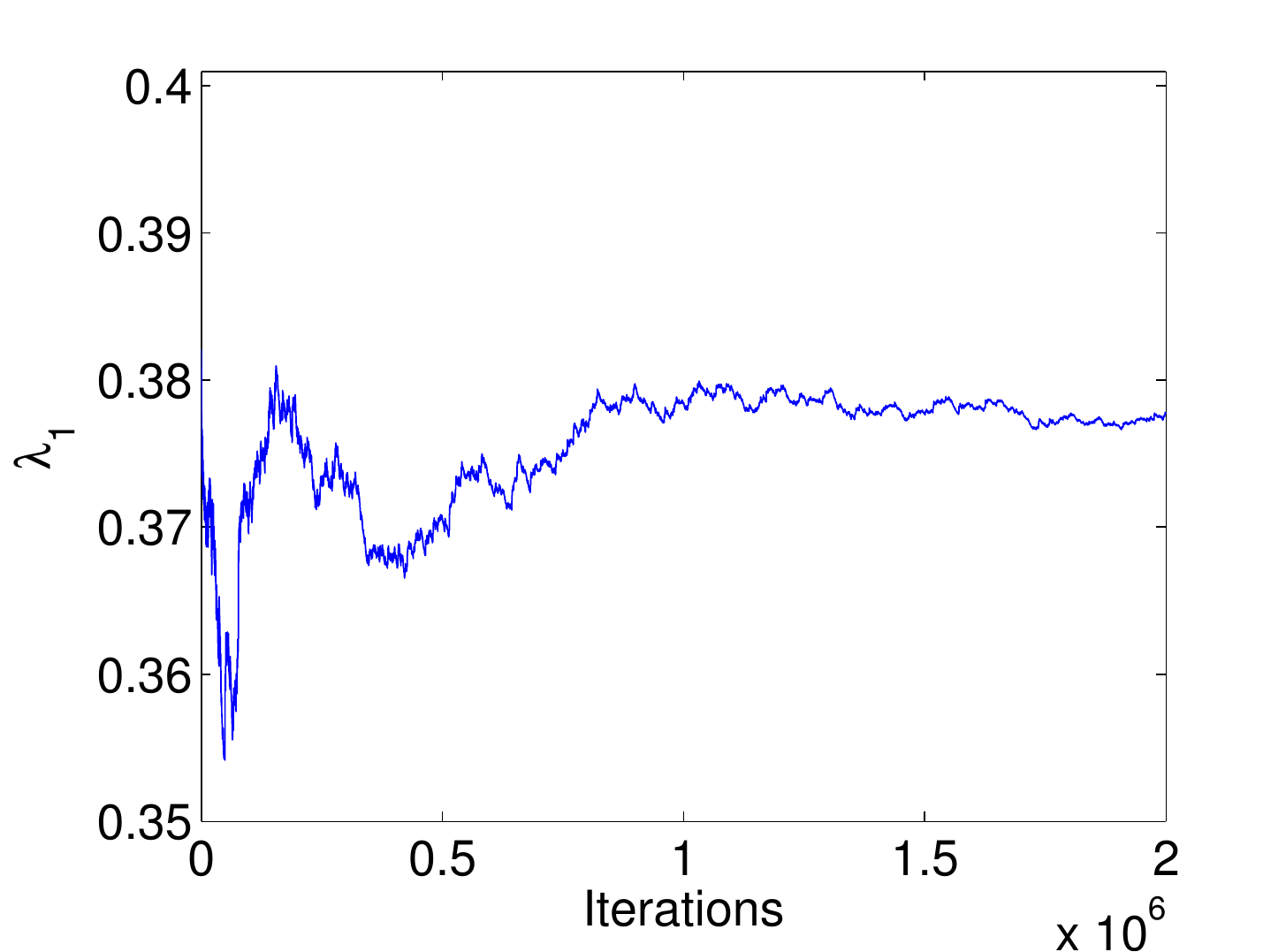}
\includegraphics[width=.49\textwidth]{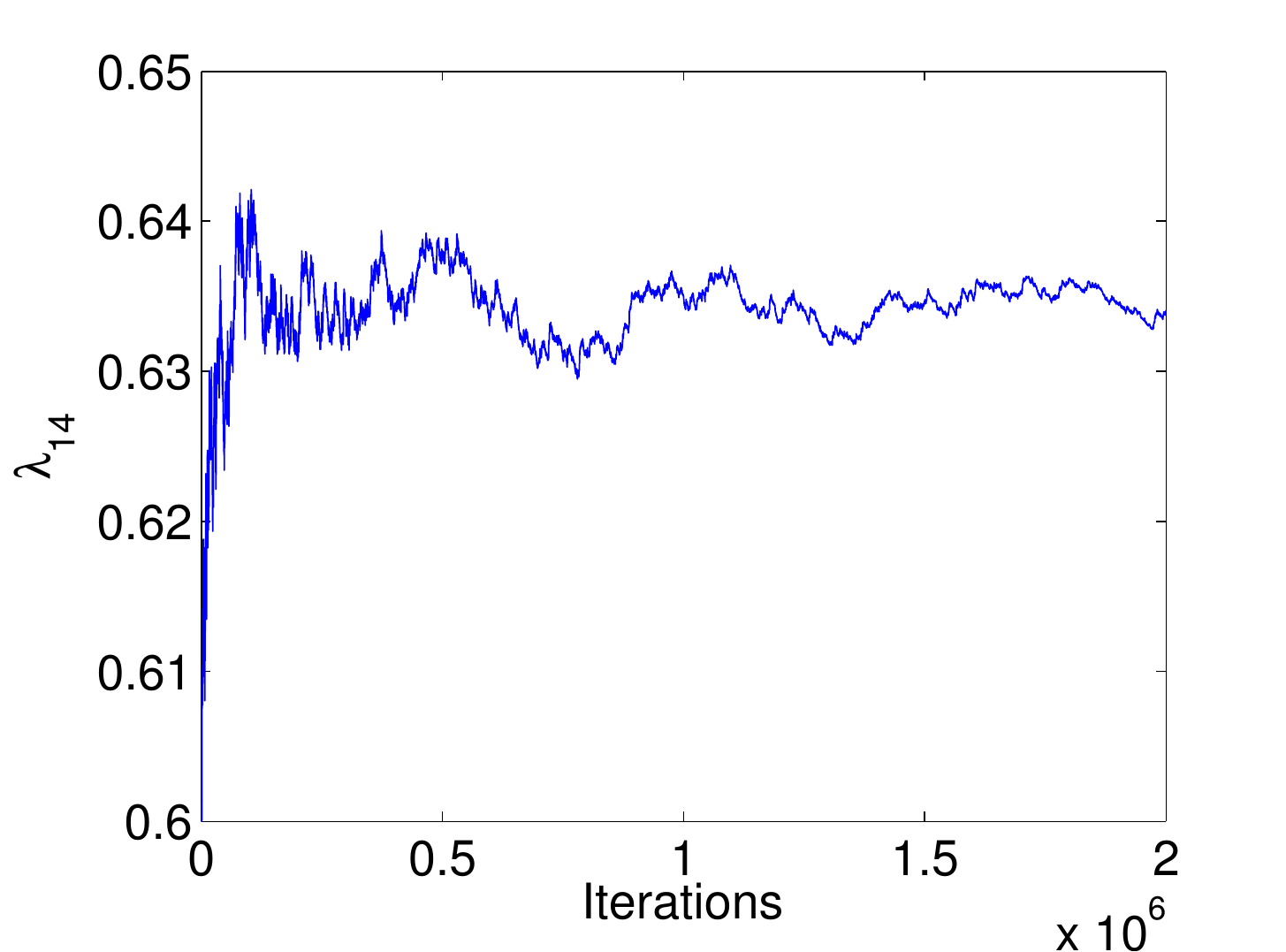}}
\caption{(for the ODE example) The eigenvalues $\lambda_1$ (Left) and $\lambda_{14}$ (Right) plotted against the number of iterations.}
\label{f:eigs}
\end{figure}

\begin{figure}
\centerline{\includegraphics[width=.49\textwidth]{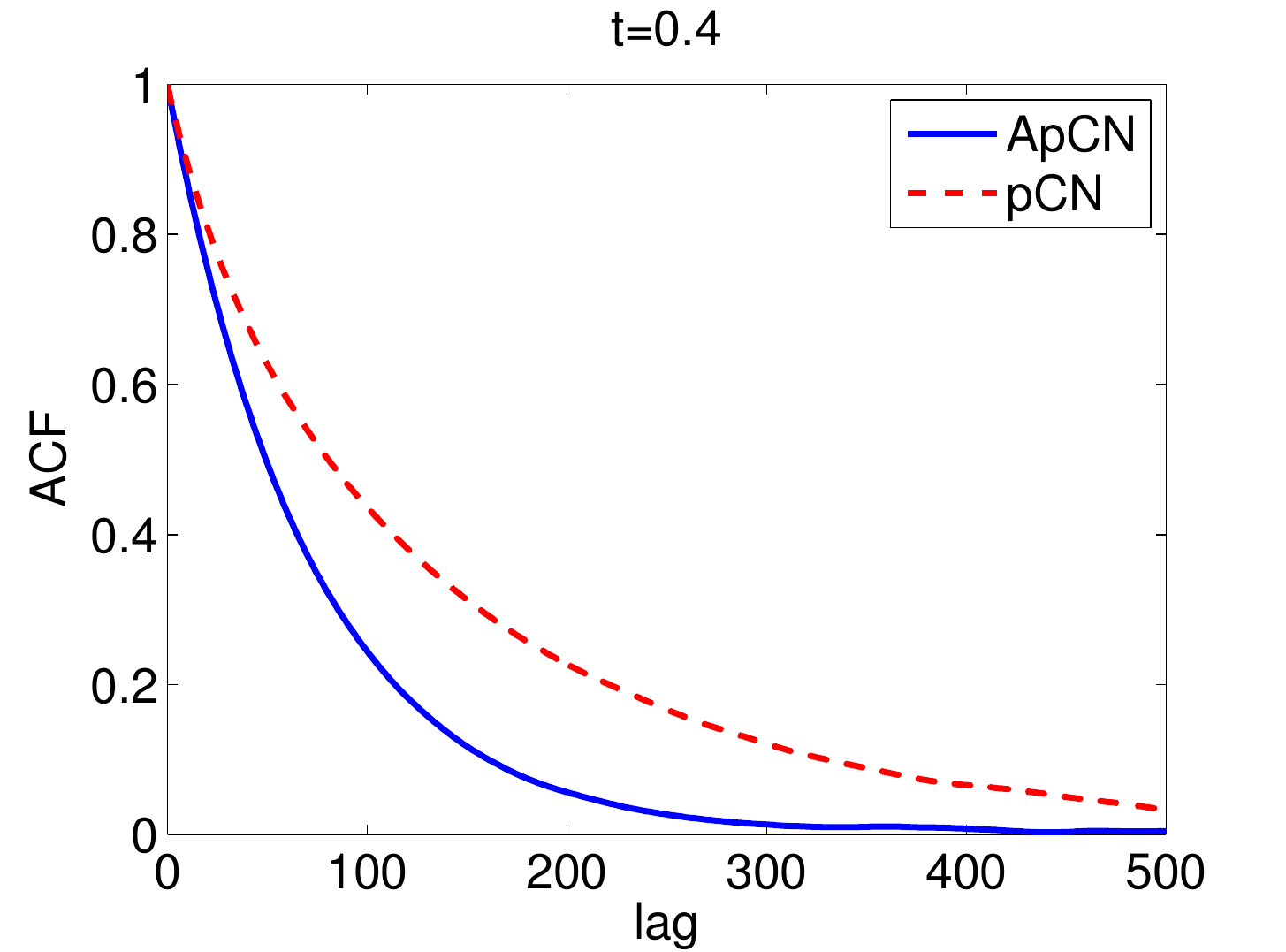}
\includegraphics[width=.49\textwidth]{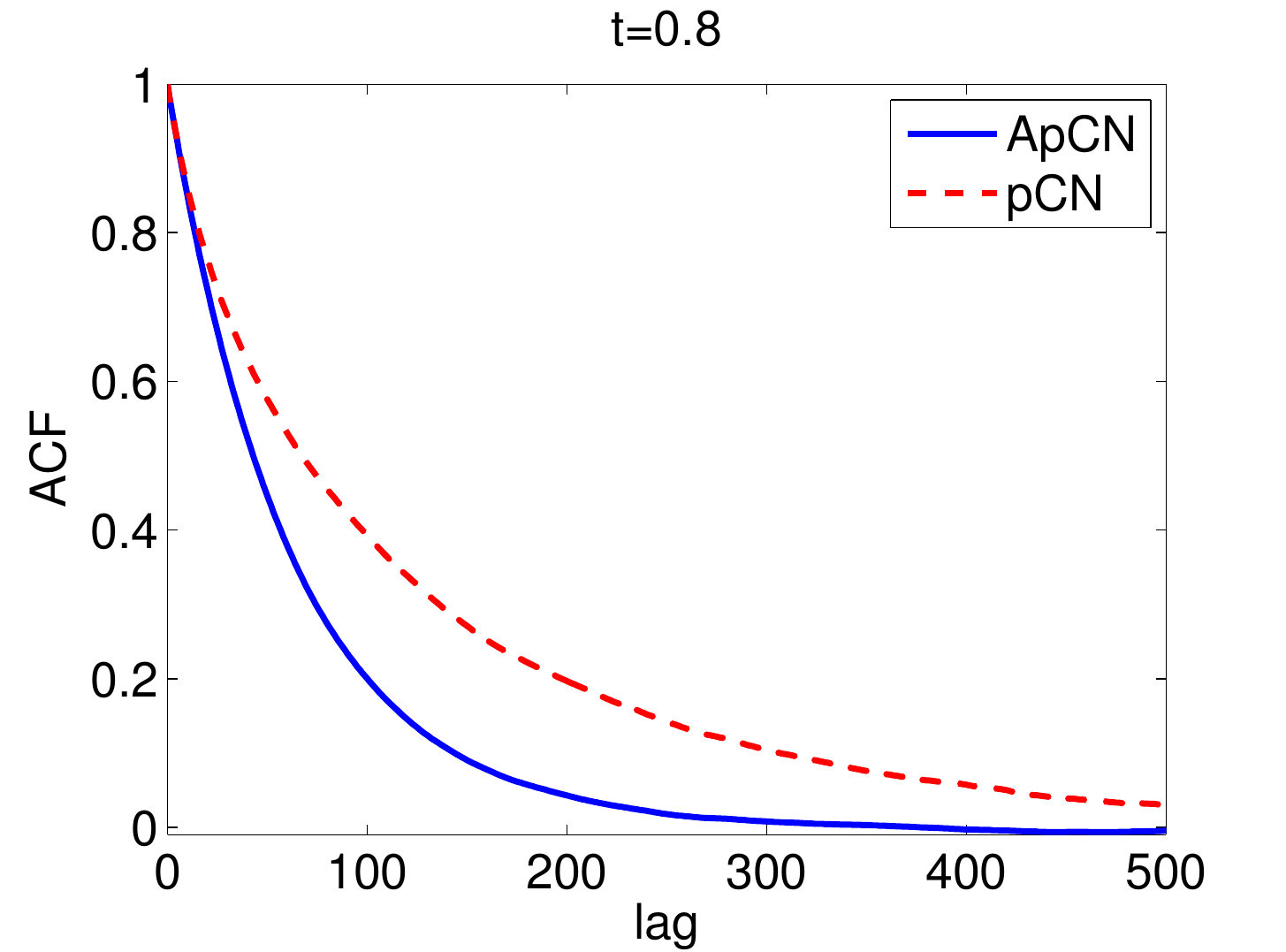}}
\caption{(for the ODE example) Autocorrelation functions (ACF) for the pCN and the ApCN methods at $t=0.2$ and $t=0.8$.}
\label{f:acf_ode}
\end{figure}

\begin{figure}
\centerline{\includegraphics[width=.49\textwidth]{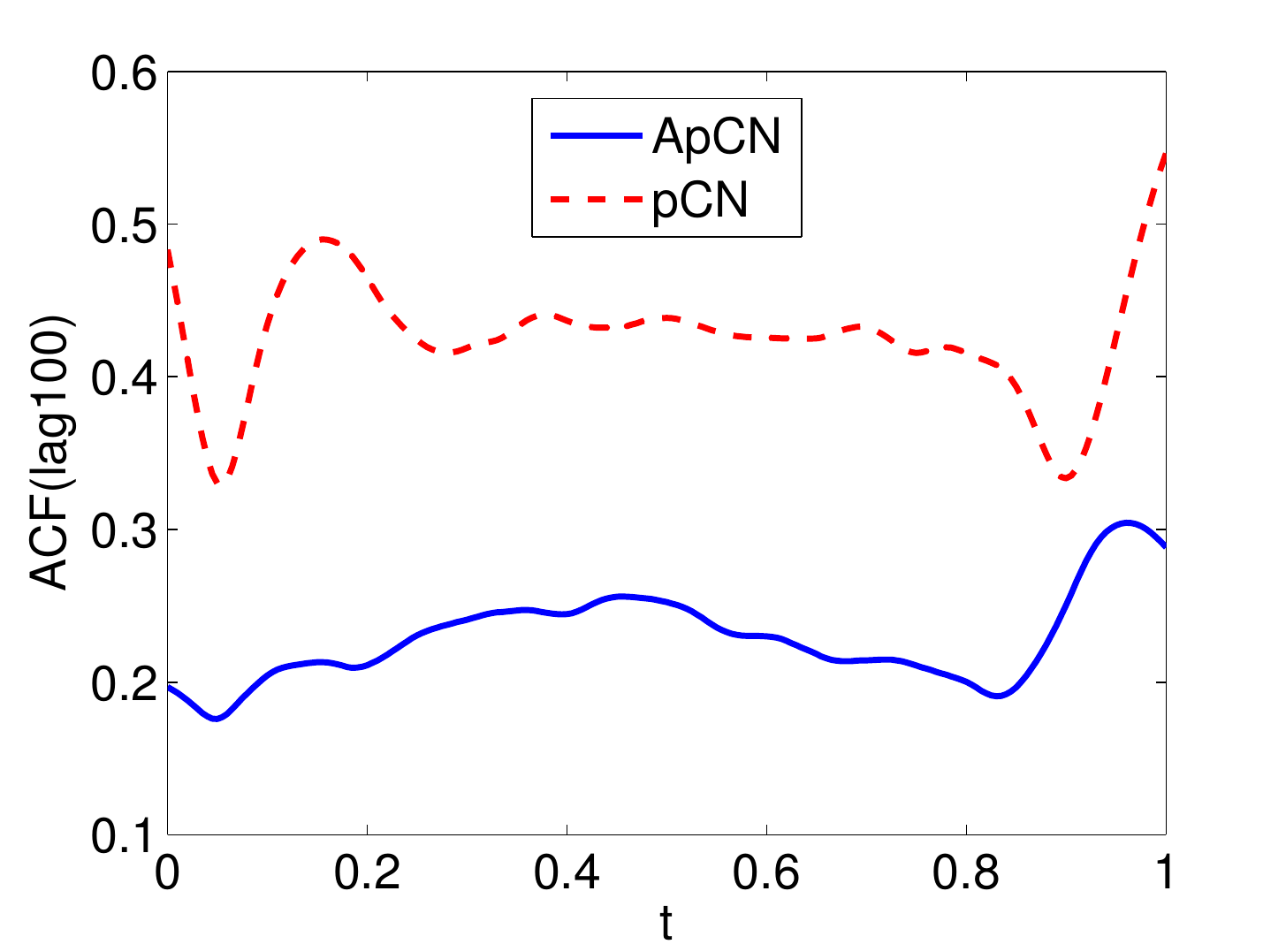}
\includegraphics[width=.49\textwidth]{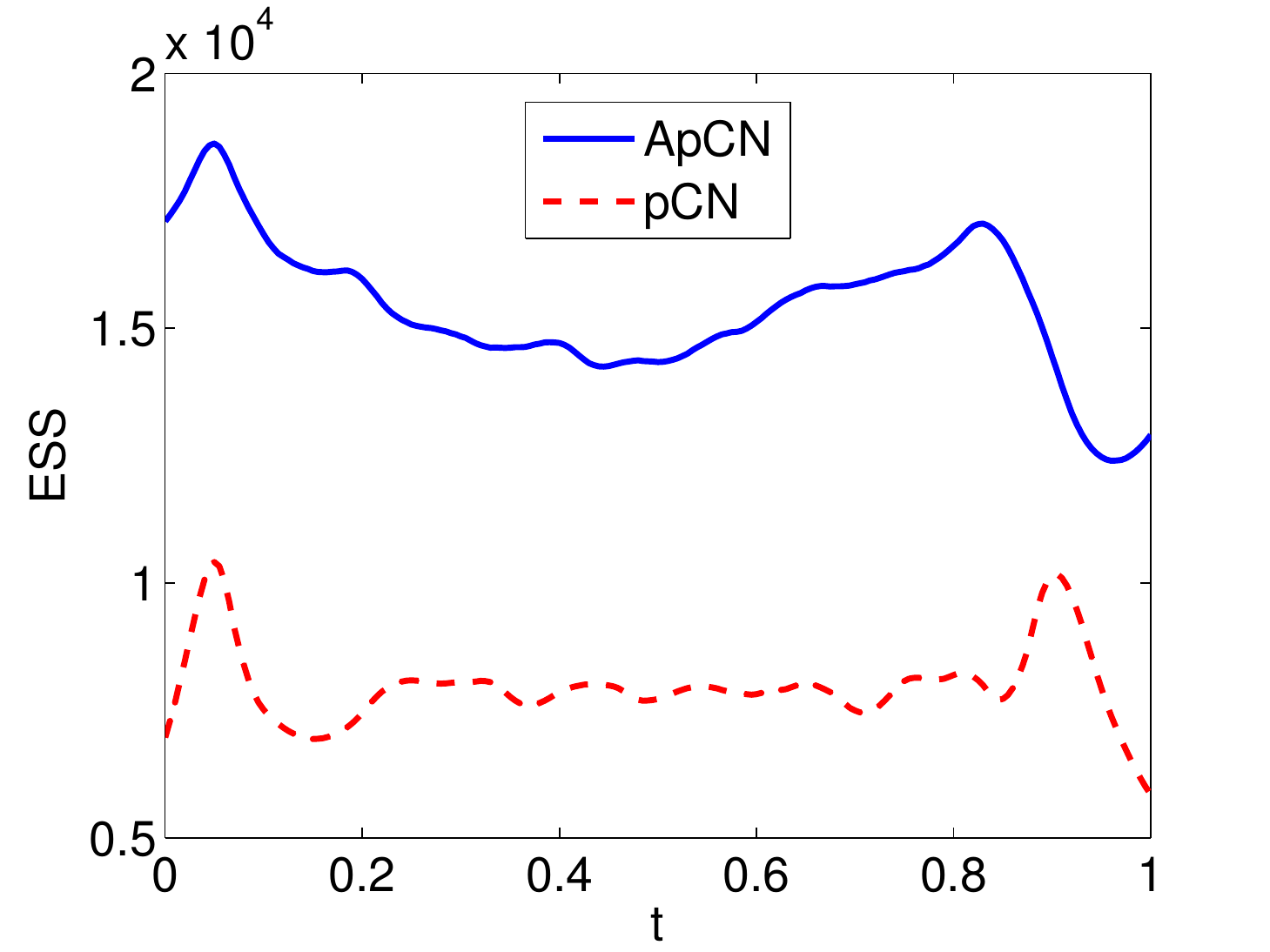}}
\caption{(for the ODE example) Left: ACF (lag 100) at each grid point. Right: ESS at each grid point.}
\label{f:acf1000-ess-ode}
\end{figure}

We show the simulation results in Figs.~\ref{f:mean_ode}: in the left figure, we show 10 randomly chosen MCMC samples from the posterior, and in the right figure,
we plot the posterior mean, as well as the $95\%$ confidence interval, both computed with the MCMC samples. 
To illustrate the diminishing of the adaption, we plot the 1st and the 14th eigenvalues against the iterations in Fig.~\ref{f:eigs}, and the plots indicate that
both parameters tend to converge to certain fixed values as the iterations proceed. 
For comparison, we also draw $1.05\times10^6$ samples from the posterior with a standard pCN algorithm where the step size is again taken to be $\beta=1/5$. 
In Figs.~\ref{f:acf_ode}, we compare the autocorrelation function (ACF) of the samples drawn by the two methods at $t=0.4$ (left) and $t=0.8$ (right),
and the ACF results show that the adaptive pCN method performs better than standard pCN.
We then compute the ACF of lag $1000$ at all the grid points, and show the results in Fig.~\ref{f:acf1000-ess-ode} (left),  
and we can see that, the ACF of the chain generated by the ApCN is clearly lower than that of the standard pCN at all the grid points. 
The effective sample size (ESS) is another popular measure of the sampling efficiency of MCMC~\cite{Kass1998}. 
ESS is computed 
by \[\mathrm{ESS} = \frac{N}{1+2\tau},\]
where $\tau$ is the integrated autocorrelation time and $N$ is the total sample size, and it gives an estimate of the number of effectively independent draws in the chain.
We compute the ESS of the unknown $u$ at each grid point and show the results in Fig.~\ref{f:acf1000-ess-ode} (right).
The results show that the ApCN algorithm  produces much more effectively independent samples than the standard pCN. 

\subsection{Estimating the Robin coefficient}
In the second example, we consider a one-dimensional heat conduction equation in the region $x\in [0,L]$ ,
\begin{subequations}
\label{e:heat}
\begin{align}
&\frac{\partial u}{\partial t}(x,t) = \frac{\partial^2 u}{\partial x^2}(x,t), \\
&u(x,0)=g(x), 
\end{align}
with the following Robin boundary conditions:
\begin{align}
&-\frac{\partial u}{\partial x}(0,t) + \rho(t) u(0,t) = h_0(t),\\
 &-\frac{\partial u}{\partial x}(L,t) + \rho(t) u(L,t) = h_1(t).
\end{align}
\end{subequations}
Suppose the functions $g(x)$,  $h_0(x)$ and $h_1(x)$ are all known, and we want to estimate the unknown Robin coefficient $\rho(t)$
from certain measurements of the temperature $u(x,t)$.  This example is studied in \cite{yao2015tv}.
Here we choose $L=1$, $T=1$ and the functions to be
\[g(x)=x^2+1,\quad  h_0 =t(2t+1),\quad h_1=2+t(2t+2).\]
 The solution is measured every $T/200$ time unit from $0$ to $T$ and the error in each measurement is assumed to be an independent Gaussian $N(0,0.1^2)$.
  In the computation, $501$ equally spaced grid points are used to represent the unknown. 
 Moreover, the prior is the same as that used in the ODE example.

\begin{figure}
\centerline{\includegraphics[width=.49\textwidth]{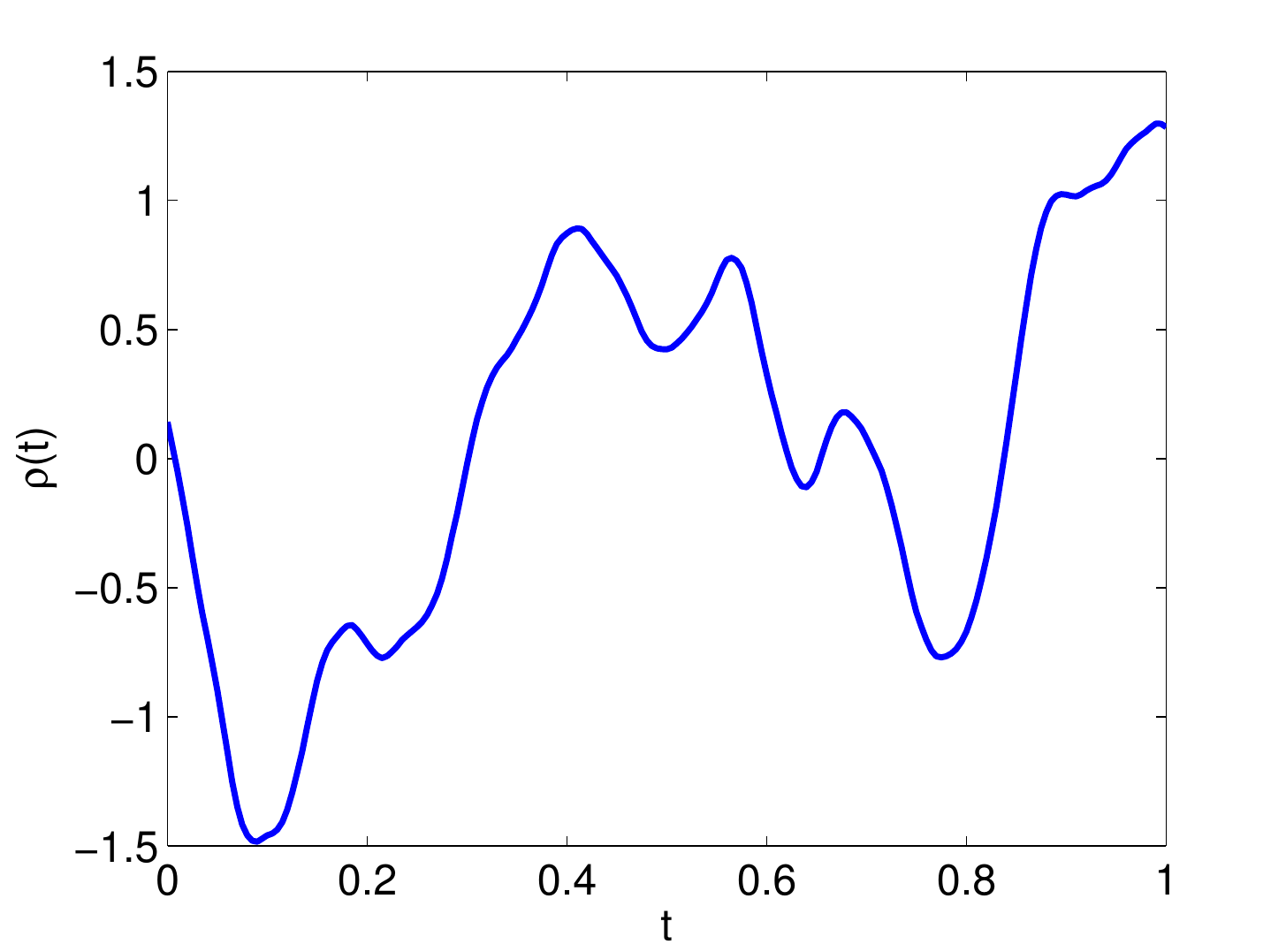}
\includegraphics[width=.49\textwidth]{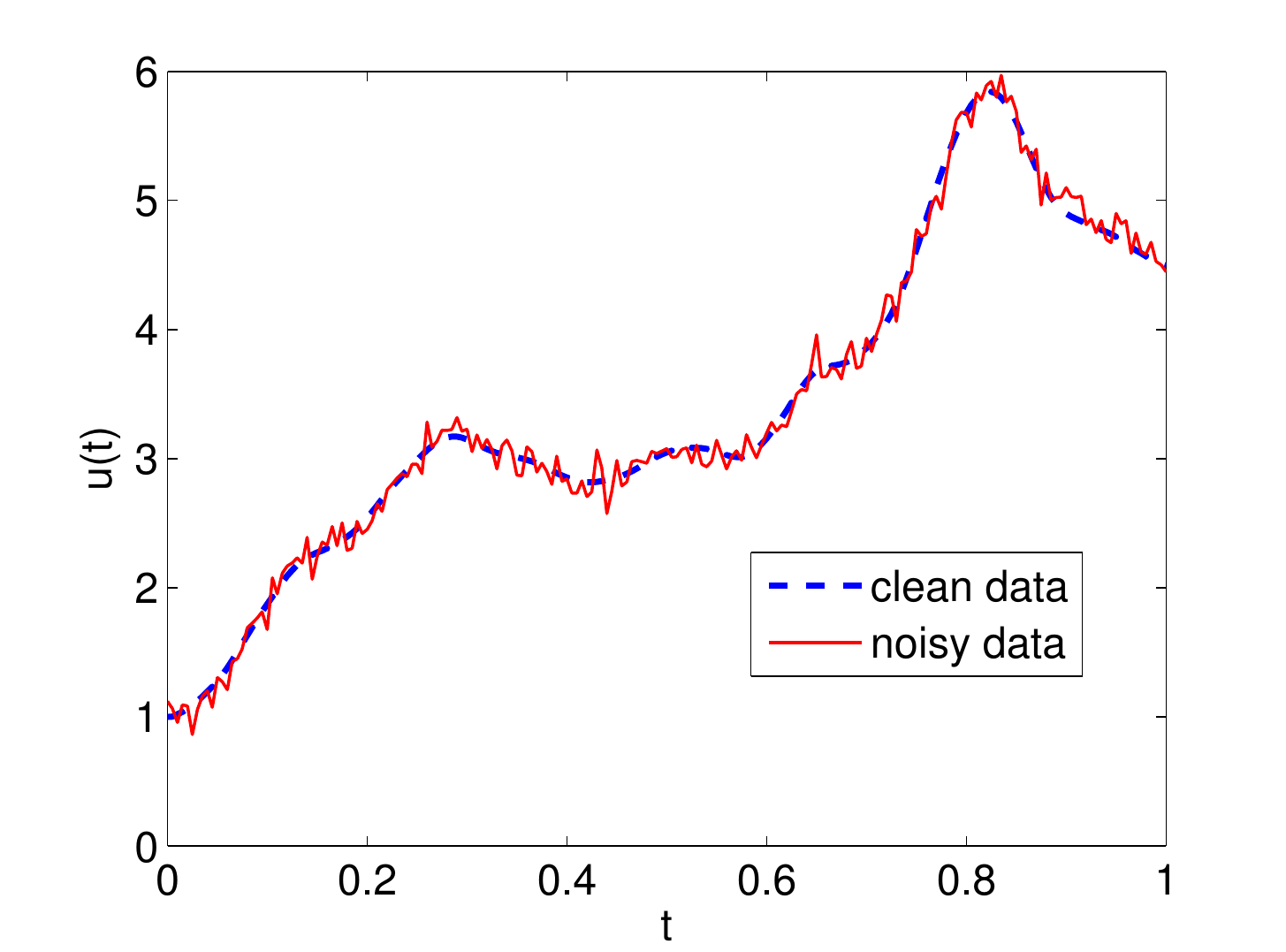}}
\caption{(for the Robin example) The truth (Left) and the data simulated with it (Right).}
\label{f:data_pde}
\end{figure}

\begin{figure}
\centerline{\includegraphics[width=.49\textwidth]{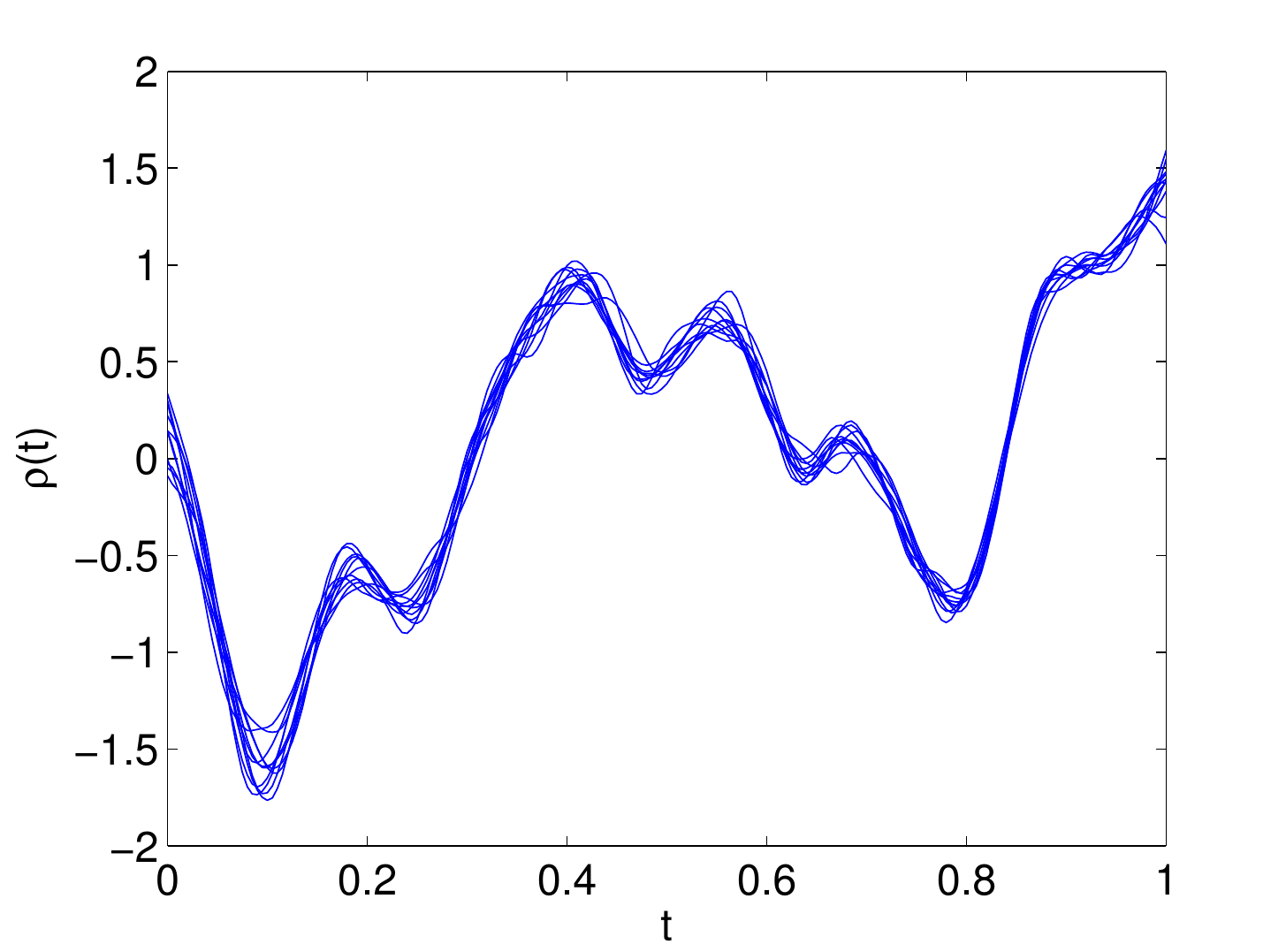}
\includegraphics[width=.49\textwidth]{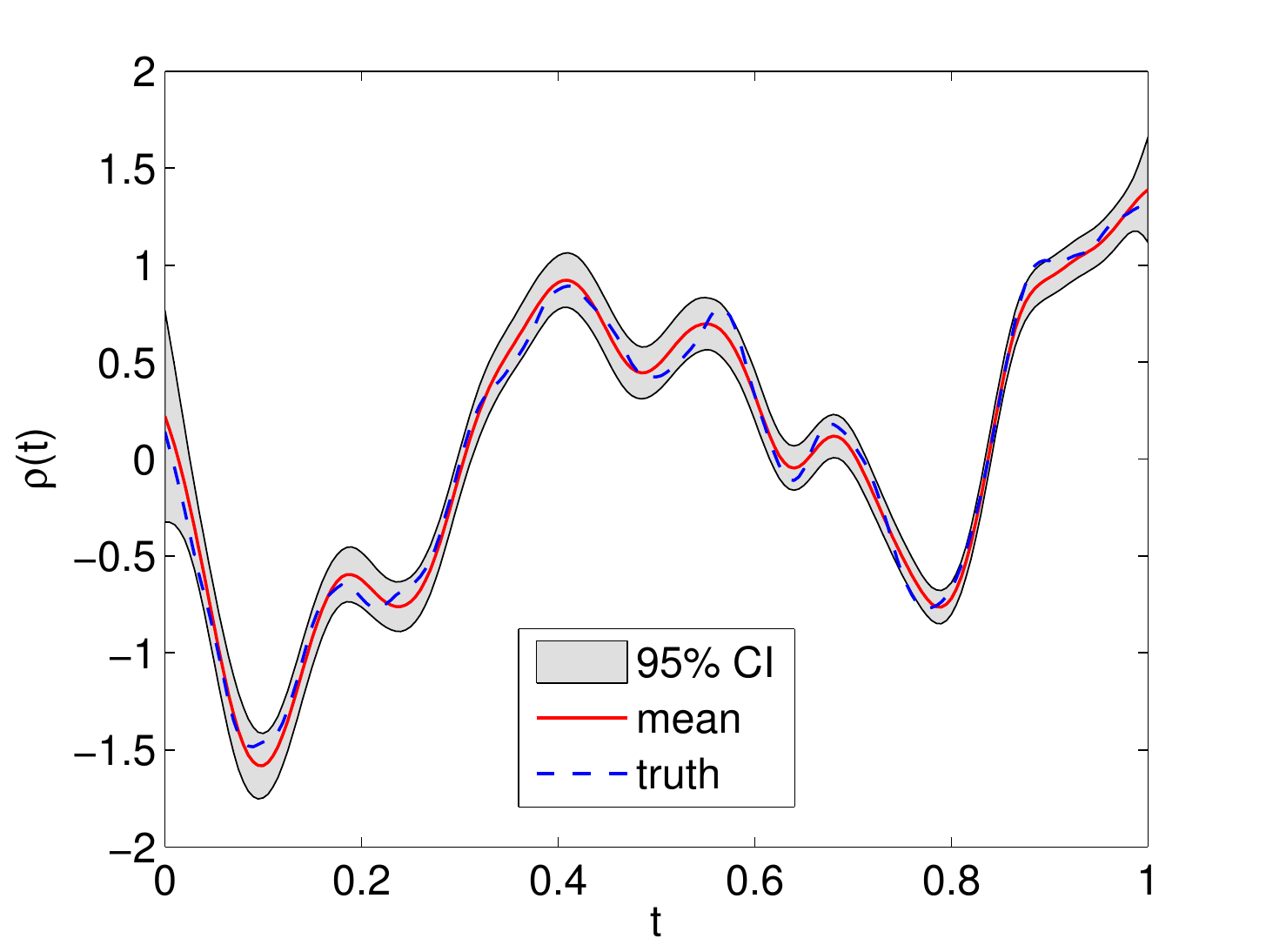}}
\caption{(for the Robin example) Left: 10 randomly drawn samples from the posterior. Right: the posterior mean and the $95\%$ confidence interval.}
\label{f:mean_robin}
\end{figure}

\begin{figure}
\centerline{\includegraphics[width=.49\textwidth]{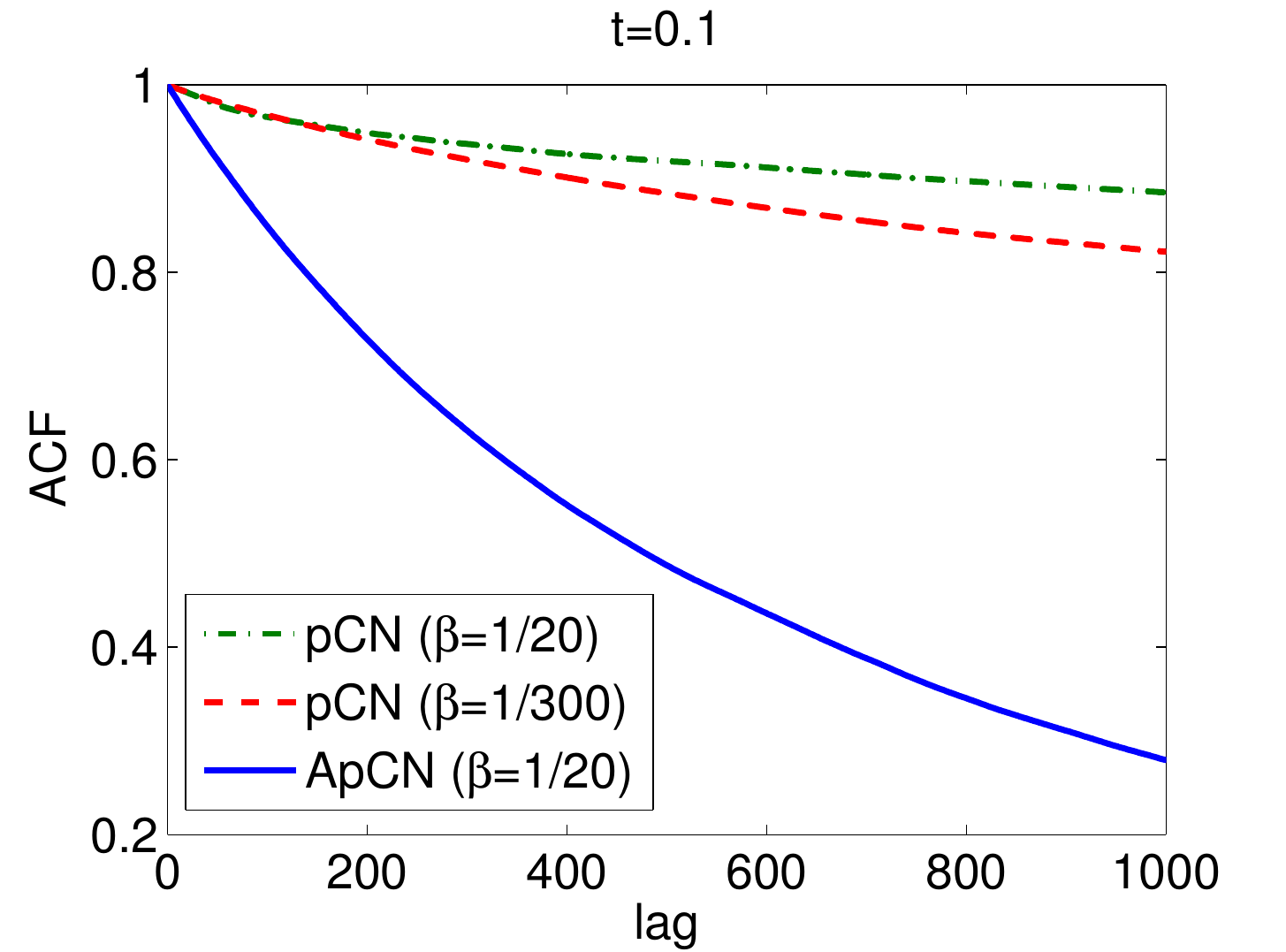}
\includegraphics[width=.49\textwidth]{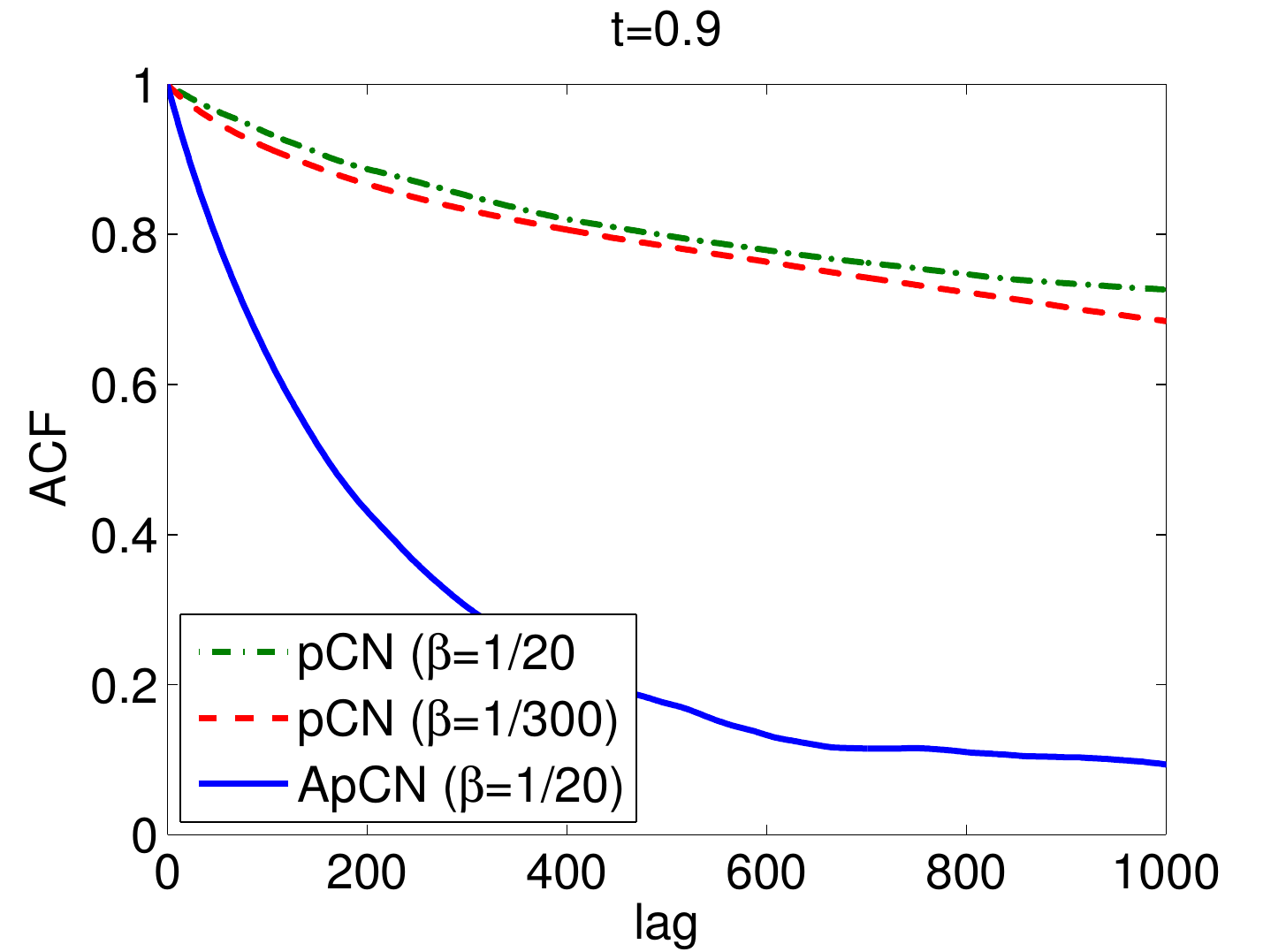}}
\caption{(for the Robin example) ACF for the pCN and the ApCN methods at $t=0.1$ and $t=0.9$.}
\label{f:acf_pde}
\end{figure}

\begin{figure}
\centerline{\includegraphics[width=.49\textwidth]{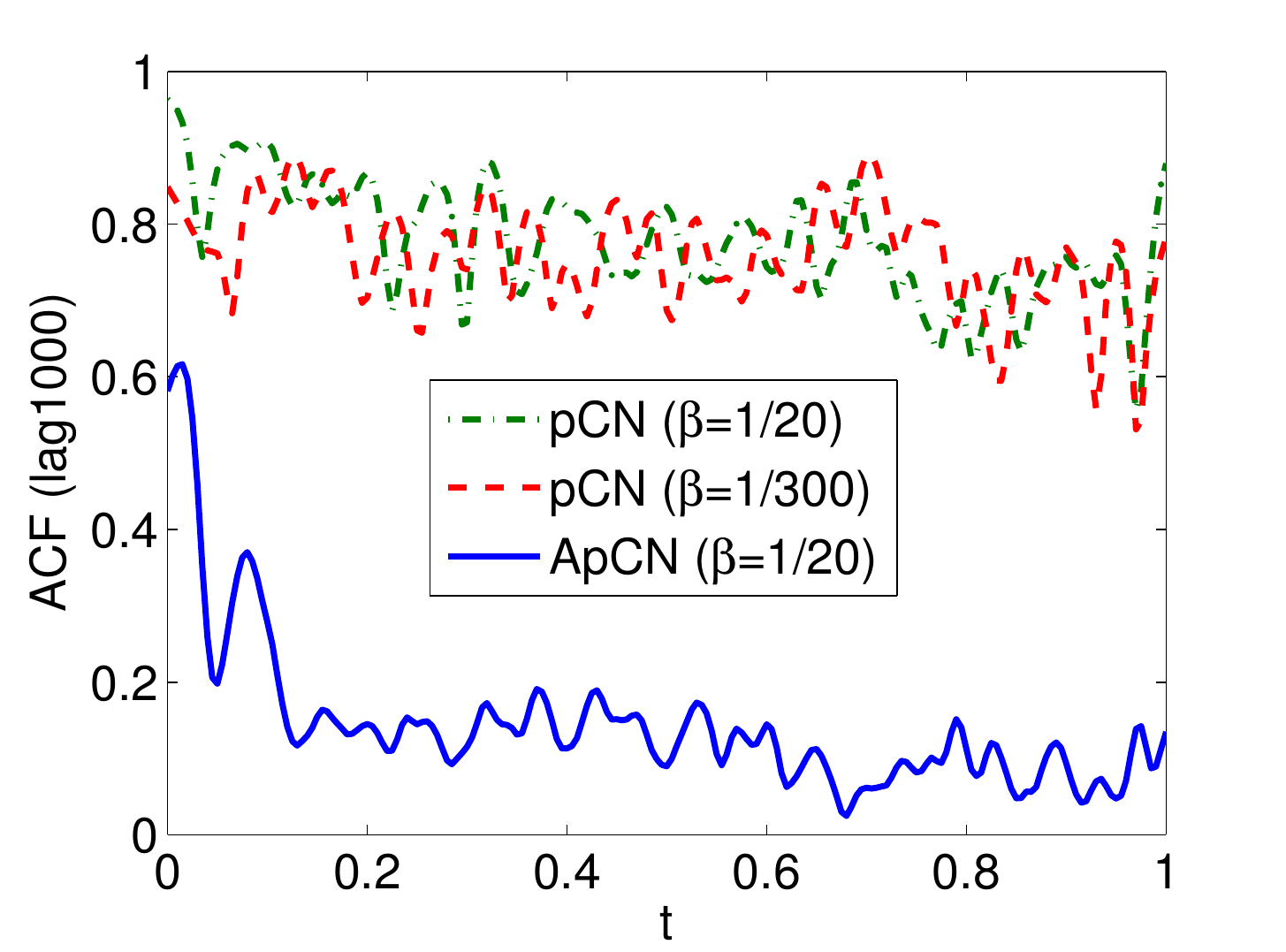}
\includegraphics[width=.49\textwidth]{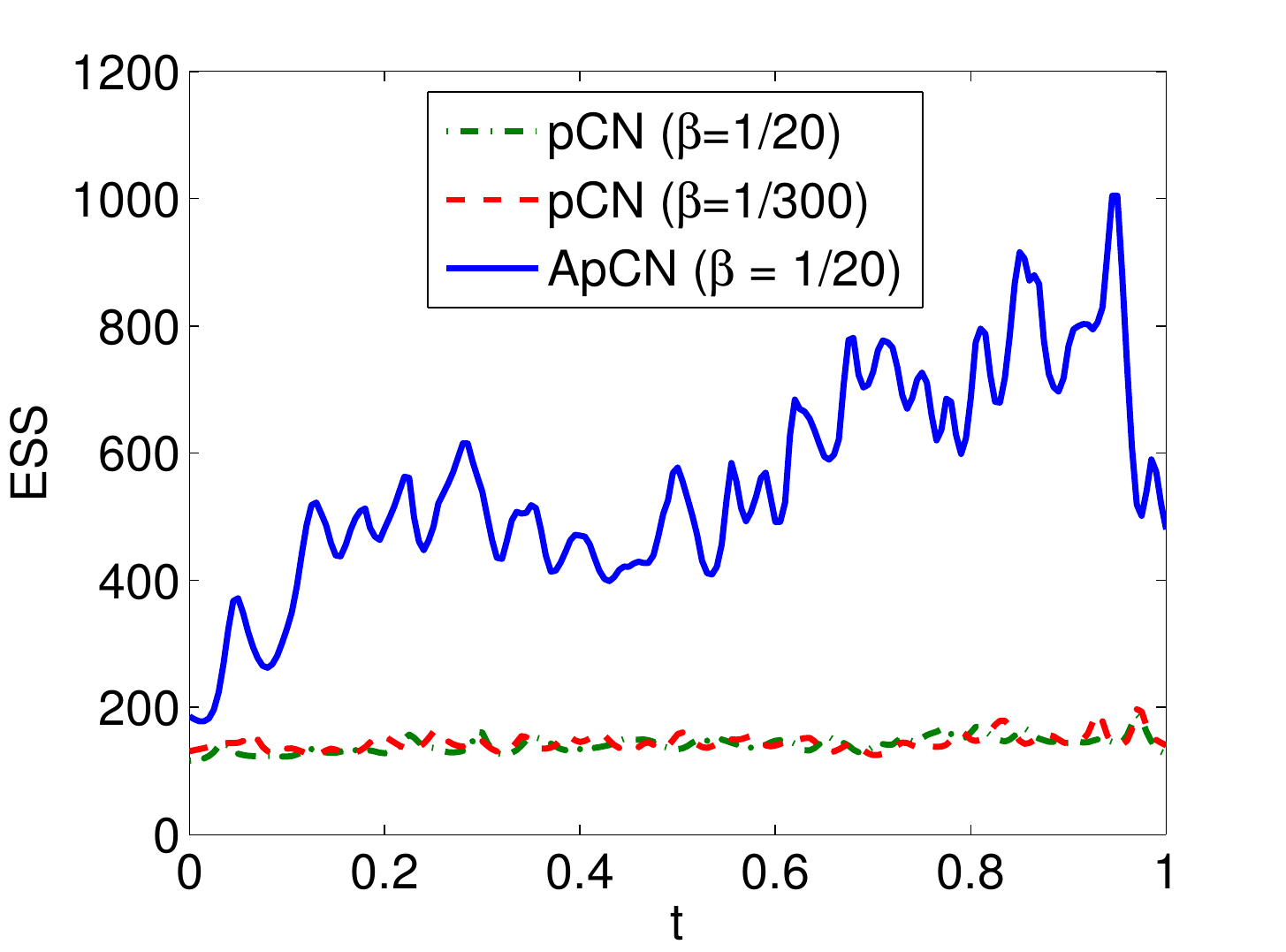}}
\caption{(for the Robin example) Left: ACF (lag 1000) at each grid point. Right: the ESS at each grid point. }
\label{f:acf1000-ess-pde}
\end{figure}

The data is generated the same as the first example, with the true Robin coefficient randomly drawn from the prior distribution. 
Both the truth and the simulated data are shown in Fig.~\ref{f:data_pde}.
 We implement the adaptive pCN algorithm, 
where we choose $\beta = 1/5$, and  $\rho=0.99$ resulting in $J=14$.  
With the algorithm, we draw $5.5\times10^5$ samples from the posterior, including $5\times10^4$ pCN samples in the pre-run,
and the average acceptance probability is around $20\%$. 
In the left plot of Figs.~\ref{f:mean_robin}, we show 10 randomly chosen MCMC samples from the posterior, and in the right plot,
we show the posterior mean and the $95\%$ confidence interval, both computed with the MCMC samples. 
Once again, we sample the posterior with standard pCN algorithm for comparison, 
and in particular we run pCN with two different stepsizes: first we use $\beta=1/5$ which is the same as that used the ApCN algorithm; 
we then use $\beta=1/300$, yielding higher acceptance probability. 
In each case, we draw $5.5\times 10^5$ samples,
and the average acceptance probability for $\beta=1/5$ is around $0.3\%$, and that for $\beta=1/300$ is around $20\%$, which matches that of the ApCN algorithm.  
In Figs.~\ref{f:acf_pde}, we compare the ACF of the samples drawn by the two methods at $t=0.1$ (left) and $t=0.9$ (right).
One can see from the figures that, the ACF of the chain generated by the pCN with $\beta=1/300$ decreases slightly faster than that with $\beta=1/20$, thanks to the higher acceptance probability, 
while the result of the ApCN is significantly better than both of them. 
We then compute the ACF of lag $1000$ as well as the ESS at all the grid points, and show the results in Figs.~\ref{f:acf1000-ess-pde}. 
Once again, both the ACF and the ESS results suggest that
the sampling efficiency of the ApCN is significantly higher than that of the standard pCN algorithm. 

\section{Conclusions}\label{sec:conclusion}
In summary, we consider MCMC simulations for Bayesian inferences in function spaces. 
In particular, we develop an adaptive variant of the pCN algorithm to improve the sampling efficiency. 
The implementation of the ApCN algorithm is rather simple, without requiring any information of the underlying models, 
and during the iteration the proposal can be efficiently updated with explicit formulas.  
We also show that the adaptive pCN algorithm satisfies certain ergodicity condition.
Finally we demonstrate the effectiveness and efficiency of the ApCN algorithm with several numerical examples. 
We expect the algorithm can be of use in many practical problems, especially in those involving blackbox models. 

Finally we note that, a major limitation of the ApCN algorithm (and of the standard pCN as well) is that the stepsize $\beta$ is ultimately restricted to be less than $1$,
while in many problems, larger step-sizes may be needed so that the resulting acceptance probability is in a favorable range. 
As a result, the ApCN algorithm in its present form may yield undesirably high acceptance probability. 
We plan to address the issue by making improvements on the present algorithm in a future work.  

\appendix

\section{Proof of Theorem~\ref{th:da}} \label{sec:proof}
We provide a proof of Theorem 2 in this appendix, which largely follows the proof for the finite dimensional adaptive Metropolis algorithm given in \cite{haario2001adaptive}.
We start with the following inequality:
\[
\begin{array}{ll}
&|Q_{n,\zeta_{n-2}}(u,A)-Q_{n+1,\zeta_{n-1}}(u,A)|\\
&=|\int_A a(u,v)q_{n,\zeta_{n-2}}(u,dv)+\delta_A(u)(1-\int_{X}{q_{n,\zeta_{n-2}}(u,dz)a(u,z)})\\
&-\int_A a(u,v)q_{n+1,\zeta_{n-1}}(u,dv)+\delta_A(u)(1-\int_{X}{q_{n+1,\zeta_{n-1}}(u,dz)a(u,z)})|\\
&\leq 2\int_X a(u,v)|\frac{dq_{n,\zeta_{n-2}}(u,\cdot)}{d\mu_0}(v)-\frac{dq_{n+1,\zeta_{n-1}}(u,\cdot)}{d\mu_0}(v)|\mu_0(dv)\\
&\leq 2\int_X |\frac{dq_{n,\zeta_{n-2}}(u,\cdot)}{d\mu_0}(v)-\frac{dq_{n+1,\zeta_{n-1}}(u,\cdot)}{d\mu_0}(v)|\mu_0(dv)\\
&\leq 2\int_X |\frac{dq_{n,\zeta_{n-2}}(u,\cdot)}{d\mu_0}(v)-\frac{d\widetilde{q}}{d\mu_0}(v)|\mu_0(dv)+2\int_X |\frac{d\widetilde{q}}{d\mu_0}(v)-\frac{dq_{n+1,\zeta_{n-1}}(u,\cdot)}{d\mu_0}(v)|\mu_0(dv),
\end{array}
\]
where $\widetilde{q}$ is the Gaussian measure that has the same mean with $q_{n,\zeta_{n-2}}(u,\cdot)$ and has the same covariance operator with $q_{n+1,\zeta_{n-1}}(u,\cdot)$.
It should be clear that $\widetilde{q}$ is equivalent to $\mu_0$.
Now let
\begin{equation}
I_1=\int_X |\frac{dq_{n,\zeta_{n-2}}(u,\cdot)}{d\mu_0}(v)-\frac{d\widetilde{q}}{d\mu_0}(v)|\mu_0(dv)
\end{equation}
and
\[
I_2=\int_X |\frac{d\widetilde{q}}{d\mu_0}(v)-\frac{dq_{n+1,\zeta_{n-1}}(u,\cdot)}{d\mu_0}(v)|\mu_0(dv).
\]
First we consider $I_1$.
Since $q_{n,\zeta_{n-2}}(u,\cdot)$ and $\widetilde{q}$ are both Gaussian measures with same mean, and their covariance operators have the same eigenfunctions and at most $J$ different eigenvalues, 
we can show that, 
\begin{equation}
I_1 =\int_{\R^J}|\prod_{i=1}^J\frac{1}{\sqrt{2\pi\beta^2\lambda_{n,i}}}\exp(-\frac{x_i^2}{2\beta^2\lambda_{n,i}})-\prod_{i=1}^J\frac{1}{\sqrt{2\pi\beta^2\lambda_{n+1,i}}}\exp(-\frac{x_i^2}{2\beta^2\lambda_{n+1,i}})|dx_1\cdots dx_J.\label{e:I1}
\end{equation}
Thanks to the modified likelihood function~\eqref{e:lh_mod}, it is easy to see that
 there exist constants $C_1,\,C_2>0$ such that
 \begin{equation}
|\lambda_{n,i}-\lambda_{n+1,i}|\leq \frac{C_1}{n}, \quad\mathrm{and}\quad \lambda_{n+1,i}\geq C_2, \label{e:c1c2}
\end{equation} 
for $i=1...J$.
Using these results, and by some elementary calculus, one can derive that $I_1< C_3/n$
for some constant $C_3>0$. 

We now consider $I_2$. Let
\[
\Delta m=(\@I-\beta^2 \@B_{n,\zeta_{n-2}}(u)\@L)^{\frac12}u-(\@I-\beta^2 \@B_{n+1,\zeta_{n-1}}(u)\@L)^{\frac12}u.
\]
and it can be seen that $\<\Delta m,\,e_i\>=0$ for $\forall i>J$. 
We re-write $I_2$ as 
\[
I_2=\int_X |1-\frac{dq_{n+1,\zeta_{n-1}}(u,\cdot)}{d\widetilde{q}}(v)|\widetilde{q}(dv),
\]
where 
\[
\frac{dq_{n+1,\zeta_{n-1}}(u,\cdot)}{d\widetilde{q}}(v)=\exp(-\frac12\Vert(\beta^2 \@B_{n,\zeta_{n-2}}(u))^{-\frac12}\Delta m\Vert^2+\langle v,(\beta^2 \@B_{n,\zeta_{n-2}}(u))^{-1}\Delta m\rangle).
\]
Similar to $I_1$, we can also write $I_2$ as a finite dimensional integral: 
\begin{multline*}
I_2=\int_{\R^J} |\prod_{i=1}^J\frac{1}{\sqrt{2\pi\beta^2\lambda_{n+1,i}}}\exp(-\frac{(x_i-\langle\Delta m,e_i\rangle)^2}{2\beta^2\lambda_{n+1,i}}\\
-\prod_{i=1}^J\frac{1}{\sqrt{2\pi\beta^2\lambda_{n+1,i}}}\exp(-\frac{x_i^2}{2\beta^2\lambda_{n+1,i}})|dx_1\cdots dx_J.
\end{multline*}
It then follows from Eqs.~\eqref{e:sigma} and \eqref{e:c1c2} that there exist constants $C_4,\,C_5>0$ such that 
\begin{equation}
|\sigma_{n,i}-\sigma_{n+1,i}|\leq \frac{C_4}{n},
\quad\mathrm{and}\quad\sigma_{n,i},\sigma_{n+1,i}\geq C_5, \label{e:c4c5}
\end{equation}
 for $i=1...J$. 
We thus have, 
\[
|\langle \Delta m,e_i\rangle|=|\sigma_{n,i}-\sigma_{n+1,i}|\cdot|\langle u,e_i\rangle|\leq\frac{C_6}{n},
\]
for some constant $C_6>0$. 
Once again, by some elementary calculus, we can obtain $I_2\leq \frac{C_7}{n}$ for some constant $C_7>0$,
which completes the proof.  

\section*{Acknowledgment}
The work was partially supported by the National Natural Science Foundation of China under grant number 11301337. 
ZH and ZY contribute equally to the work.

\bibliographystyle{plain}
\bibliography{apcn}

\end{document}